\documentclass[aip,pop,reprint,amsmath,amssymb,citeautoscript]{revtex4-2}
\pdfoutput=1

\usepackage{bm}
\usepackage{graphicx}
\usepackage{comment}

\usepackage{fancyhdr}

\usepackage{hyperref}
\hypersetup{colorlinks=true, urlcolor=cyan, linkcolor=cyan, citecolor=cyan}

\begin{document}


\title{Linear embedding of nonlinear dynamical systems and prospects for efficient quantum algorithms}


\author{Alexander Engel}
\email{alen3220@colorado.edu}
\author{Graeme Smith}
\altaffiliation{Also at JILA, University of Colorado, Boulder, Colorado 80309, USA.}
\author{Scott E. Parker}
\affiliation{Department of Physics, University of Colorado, Boulder, Colorado 80309, USA}



\begin{abstract}
The simulation of large nonlinear dynamical systems, including systems generated by discretization of hyperbolic partial differential equations, can be computationally demanding. Such systems are important in both fluid and kinetic computational plasma physics. This motivates exploring whether a future error-corrected quantum computer could perform these simulations more efficiently than any classical computer. We describe a method for mapping any finite nonlinear dynamical system to an infinite linear dynamical system (embedding) and detail three specific cases of this method that correspond to previously-studied mappings. Then we explore an approach for approximating the resulting infinite linear system with finite linear systems (truncation). Using a number of qubits only logarithmic in the number of variables of the nonlinear system, a quantum computer could simulate truncated systems to approximate output quantities if the nonlinearity is sufficiently weak. Other aspects of the computational efficiency of the three detailed embedding strategies are also discussed.
\end{abstract}


\maketitle

\fancyhf{}
\renewcommand{\headrulewidth}{0pt}
\lfoot{\bf \footnotesize This article may be downloaded for personal use only. Any other use requires prior permission of the author and AIP Publishing. This article appeared in Physics of Plasmas 28, 062305 (2021) and may be found at \url{https://doi.org/10.1063/5.0040313}.}
\thispagestyle{fancy}

\section{Introduction}

Plasma applications have historically been at the forefront of high-performance computing, and plasma physics computations are routinely performed using the largest available supercomputers \cite{Murty1983,Service2018}. Even so, resolution, domain size, and dimensionality often limit the realism of plasma simulations. More computing power would allow for better understanding and more accurate predictions of the behavior of both natural and experimental plasmas. Quantum computers may eventually provide some of this computing power. Although they are at an early stage of development, error-corrected quantum computers have the potential to be much more powerful than classical computers, at least for some tasks\cite{Preskill2018}. Plasma simulations are typically nonlinear, and the investigation of the power of quantum computation for nonlinear simulations, including plasma simulations, is an active research area.\cite{Joseph2020, Dodin2020, Variational, QuantumFluid, Dissipative, Lloyd2020, Shi2020}

Consider a kinetic plasma physics computation composed of the following steps. First, particle distribution functions $f_s({\bf x}, {\bf v}, t)$, where $s$ is the particle species, are initialized at time $t=0$ to specified functions of position ${\bf x}$ and velocity ${\bf v}$. Second, the distribution functions are time-evolved numerically using the Vlasov equation or similar. Finally, some simple functional of $f_s({\bf x}, {\bf v}, t)$ is extracted as the output of the computation. This process may also be repeated a small number of times to obtain different output quantities or to evaluate outputs at different $t$. If all these steps can be done efficiently, many important kinetic plasma physics problems can be solved. But in practice, the number of variables required to represent $f_s({\bf x}, {\bf v}, t)$ makes this computation extremely expensive in general.

An analogous situation occurs when simulating many-particle, quantum-mechanical systems. Both the initial quantum-mechanical state and the desired output may have simple functional forms, but, since the number of variables needed to represent the state grows exponentially with the number of particles, this computation can be intractable. On the other hand, with a quantum computer, exponentially large states can be represented efficiently, and general quantum-mechanical simulations can be performed with exponentially lower computational costs than with a regular computer. This possibility is what led Feynman to originally propose the idea of quantum computation \cite{Feynman1982}.

Moreover, universal quantum computers can obtain speedups for computations that are not quantum-mechanical in origin. For example, when the Vlasov-Maxwell system is linearized about a Maxwellian background, it can be transformed to have the form of a quantum-mechanical system \cite{Engel2019}. Then a quantum algorithm, i.e., an algorithm designed to run on a quantum computer, can potentially simulate this plasma physics system with a large speedup. In particular, we previously formulated an efficient quantum algorithm for the linear Landau damping problem and discussed extensions to six-dimensional phase space and full electromagnetics \cite{Engel2019}. Others who have studied the application of quantum algorithms to classical plasma physics problems include Parker and Joseph\cite{Parker2020} and Dodin and Startsev\cite{Dodin2020}.

Quantum algorithms that perform general linear computations with a large speedup have also been designed. An important example is provided by the linear problem $A{\bf x} = {\bf b}$, where $A$ is an invertible matrix, and ${\bf b}$ is a known vector. This can be solved by a quantum linear systems algorithm (QLSA) with costs only logarithmic in $N$, where $N$ is the length of ${\bf b}$ \cite{Harrow2009, QLSA2017}. To go into a bit more detail, it is assumed that one can efficiently prepare a quantum state
\[
\vert \psi_i \rangle \propto \sum_j b_j \vert j \rangle,
\]
where the proportionality constant is determined by requiring $\vert \psi_i \rangle$ to be normalized. The index $j$ runs over the indices of ${\bf b}$, and the state $\vert j \rangle$ corresponds to a sequence of qubits with values matching the bits of the binary representation of $j$. Whether the $\vert \psi_i \rangle$ state can be prepared efficiently depends on ${\bf b}$, but it is possible in some general cases, such as if each $b_j$ can be computed efficiently and $\max_j \vert b_j \vert / \vert {\bf b} \vert = \mathcal{O}(1/\sqrt{N})$, where $\vert {\bf b} \vert=\sqrt{\sum_j b_j^* b_j}$ denotes the complex vector magnitude\cite{Soklakov2006}. The QLSA then outputs a state
\[
\vert \psi_f \rangle \propto \sum_j x_j \vert j \rangle,
\]
where ${\bf x} = A^{-1} {\bf b}$. It is also assumed that the desired output quantity can be efficiently extracted from $\vert \psi_f \rangle$. A QLSA can then obtain an exponential speedup over a classical algorithm solving $A{\bf x} = {\bf b}$ subject to conditions on the matrix $A$, including that its condition number is not large \cite{Harrow2009}.

Quantum linear systems algorithms can be used to perform various other linear computations, including the evolution of systems of linear differential equations \cite{ODE2017}. However, we are really interested in the evolution of nonlinear systems. A good example is the Vlasov-Poisson system for electrons:
\begin{align*}\frac{\partial f({\bf x}, {\bf v})}{\partial t} = -{\bf v} \cdot \bm{\nabla} f({\bf x}, {\bf v}) + \frac{e}{m_e} {\bf E}({\bf x}) \cdot \frac{\partial f({\bf x}, {\bf v})}{\partial {\bf v}},\\
{\bf E}({\bf x}) = -\bm{\nabla} \phi({\bf x}), \quad \nabla^2 \phi({\bf x}) = -\frac{\rho({\bf x})}{\epsilon_0},\\ \rho({\bf x}) = -e \int f({\bf x} , {\bf v})d{\bf v},
\end{align*}
where the $t$ dependence of each variable has been omitted for brevity. A more compact expression for this system can be obtained by replacing the electric field ${\bf E}({\bf x})$ with its value obtained through Coulomb's law:
\begin{multline}\label{eq:esv}
\frac{\partial f({\bf x}, {\bf v})}{\partial t} = -{\bf v} \cdot \bm{\nabla} f({\bf x}, {\bf v}) \\- \frac{1}{4\pi}\left(\int f({\bf x}', {\bf v}') \frac{{\bf x}-{\bf x}'}{\vert {\bf x}-{\bf x}' \vert^3} d{\bf x}' d{\bf v}'\right) \cdot \frac{\partial f({\bf x}, {\bf v})}{\partial {\bf v}},
\end{multline}
where we have also rescaled quantities so that they are all dimensionless. The Debye length $\lambda_{De}$ is the distance unit, the plasma frequency $\omega_{pe}$ is the inverse time unit, and $n_e / (\lambda_{De} \omega_{pe})^3$ is the unit in which $f({\bf x}, {\bf v})$ is expressed, where $n_e$ is the electron number density.  
To solve Eq.~(\ref{eq:esv}), $f({\bf x}, {\bf v})$ can be represented on a grid, but a very large number of grid cells will be required in general.

One way to simulate nonlinear dynamics on a quantum computer is with the Koopman-von Neumann approach introduced by Joseph\cite{Joseph2020} and studied further by Dodin and Startsev \cite{Dodin2020}. We explore a different approach, with a goal of having the quantum computational costs scale only logarithmically with the number of variables $N$ for simulation problems that have classical costs linear in $N$. For instance, this would mean costs logarithmic in the number of phase-space grid cells used to represent Eq.~(\ref{eq:esv}). But to apply quantum algorithms to a nonlinear simulation problem, we first map the nonlinear system to a linear one. The basic idea here can be understood as follows. Consider the extremely simple nonlinear system
\[
d_t x(t) = x^2(t).
\]
If we introduce a set of variables $y_r = x^r$ for $r=0,1,2,...$, then we obtain the infinite linear system
\[
d_t y_r(t) = r y_{r+1}(t).
\]
This demonstrates one way that a nonlinear system can be mapped to an infinite linear one. In Sec.~\ref{sec:embed} we consider the general problem of mapping nonlinear dynamical systems to infinite linear ones and review three specific mappings, with the one in Sec.~\ref{sec:carl} generalizing the above example. Then in Sec.~\ref{sec:trunc} we suggest a way to truncate the infinite-dimensional linear systems to obtain linear systems of size poly($N$), where $N$ is the number of variables of the original nonlinear system, although these truncated systems might only provide good approximations to output quantities if the nonlinearity is sufficiently weak. Some analysis of the efficiency of a quantum computation based on the truncated linear systems is provided in Sec.~\ref{sec:eff}, and we discuss our findings in Sec.~\ref{sec:dis}.

\section{Linear Embedding}\label{sec:embed}

First, we describe a method for expressing nonlinear dynamical systems as infinite-dimensional, linear dynamical systems. We write the original dynamical system as
\begin{equation}\label{eq:orig}
d_t {\bf z}(t) = {\bf F}[{\bf z}(t)],
\end{equation}
where ${\bf z}(t)$ is the vector of variables, and ${\bf F}({\bf z})$ is a vector function of the components of ${\bf z}$. We use $N$ to denote the length of these vectors, i.e., the number of variables. Equation (\ref{eq:orig}) is a system of ordinary differential equations; partial differential equations such as the Vlasov equation can be converted to this form through, e.g., spectral, finite element, or finite volume methods.

To map Eq.~(\ref{eq:orig}) to a linear system we define a set of states 
\begin{equation}\label{eq:zform}
\vert \psi(t) \rangle := e^{{\bf z}(t) \cdot \hat{\bf w}} \vert {\bf 0} \rangle,
\end{equation}
where $\hat{\bf w}$ is a vector of operators, and $\vert {\bf 0} \rangle$ is a fixed state. Up to an overall normalization factor, these states just generalize those considered by Kowalski \cite{BlackBook1}. Additionally, we introduce a vector of operators $\hat{\bf z}$ that satisfies
\begin{align}\label{eq:req}
\hat{\bf z} \vert {\bf 0} \rangle &= {\bf 0}, & [\hat{z}_j, \hat{w}_k] &= \delta_{jk}.
\end{align}
The reason for denoting these operators by $\hat{\bf z}$ is revealed by the following evaluation:
\begin{align}
\begin{split}
\hat{z}_j e^{{\bf z} \cdot \hat{\bf w}} \vert {\bf 0} \rangle &= \{[\hat{z}_j, e^{{\bf z} \cdot \hat{\bf w}}] + e^{{\bf z} \cdot \hat{\bf w}} \hat{z}_j\} \vert {\bf 0} \rangle \\&= \sum_{s=0}^\infty \frac{1}{s!} \left[\hat{z}_j, ({\bf z} \cdot \hat{\bf w})^s\right] \vert {\bf 0} \rangle \\&= \sum_{s=0}^\infty \frac{1}{s!} \sum_{k=0}^{s-1}({\bf z} \cdot \hat{\bf w})^{k} [\hat{z}_j, {\bf z} \cdot \hat{\bf w}] ({\bf z} \cdot \hat{\bf w})^{s-k-1} \vert {\bf 0} \rangle\\&= \sum_{s=0}^\infty z_j \frac{s}{s!} ({\bf z} \cdot \hat{\bf w})^{s-1} \vert {\bf 0} \rangle \\&= z_j e^{{\bf z} \cdot \hat{\bf w}} \vert {\bf 0} \rangle,
\end{split}
\end{align}
and therefore,
\begin{equation}\label{eq:eigz}
\hat{z}_j \vert \psi(t) \rangle = z_j(t) \vert \psi(t) \rangle.
\end{equation}
Note that, since the $\hat{\bf z}$ operators have eigenvalues of every variable value taken at any time, continuous evolution of ${\bf z}(t)$ implies that the space spanned by the $\vert \psi(t) \rangle$ states is infinite dimensional. Also, Eq.~(\ref{eq:eigz}) implies that the $\hat{\bf z}$ operators commute within this space: $\hat{z}_j \hat{z}_k \vert \psi(t) \rangle = \hat{z}_k \hat{z}_j \vert \psi(t) \rangle$ for every $\vert \psi(t) \rangle$ state and linear combinations of them. One  consequence is that the embedding does not introduce any quantum-mechanical effects, such as the uncertainty principle, into the evolution of the classical system.

Next, differentiating Eq.~(\ref{eq:zform}) with respect to time gives
\begin{align}\label{eq:check}
\begin{split}
d_t \vert \psi(t) \rangle &= d_t {\bf z}(t) \cdot \hat{\bf w} \vert \psi(t) \rangle \\&= \hat{\bf w} \cdot {\bf F}[{\bf z}(t)] \vert \psi(t) \rangle = \hat{\bf w} \cdot {\bf F}(\hat{\bf z}) \vert \psi(t) \rangle,
\end{split}
\end{align}
where we assume that all components of ${\bf F}({\bf z})$ are analytic functions. Then ${\bf F}(\hat{\bf z})$ is defined by replacement of ${\bf z}$ with $\hat{\bf z}$ in the power series representation of ${\bf F}({\bf z})$. Equation (\ref{eq:check}) shows that the $\vert \psi(t) \rangle$ states evolve linearly according to
\begin{equation}\label{eq:evolve}
d_t \vert \psi(t) \rangle = \hat{M} \vert \psi(t) \rangle,
\end{equation}
where
\begin{equation}\label{eq:M}
\hat{M} = \hat{\bf w} \cdot {\bf F}(\hat{\bf z}).
\end{equation}
The finite-time evolution can be expressed as
\begin{equation}
\vert \psi(t) \rangle = e^{\hat{M} t} \vert \psi(0) \rangle,
\end{equation}
where $\vert \psi(0) \rangle = e^{{\bf z}(0) \cdot \hat{\bf w}} \vert {\bf 0} \rangle$ depends on the initial state of the classical system, ${\bf z}(0)$.

Viewing the linear evolution as occurring in a Hilbert space containing the $\vert \psi(t) \rangle$ states, the inner product with some state $\vert c \rangle$ yields an output quantity
\begin{equation}
c(t) := \langle c \vert \psi(t) \rangle = \langle c \vert e^{\hat{M} t} \vert \psi(0) \rangle
\end{equation}
that depends on the final state of the nonlinear dynamical system. To go into more detail, we now consider a few specific instances of the embedding method. All of these have been previously studied in some form by other authors (e.g., \cite{Koopman, Neumann1, Neumann2, Carleman1932, Steeb1983, Chirikov, Alanson1992, Kow1997}), but we give an exposition of them here to illustrate various aspects of linear embedding and to collect results that will be used in later sections.

\subsection{Carleman Embedding}\label{sec:carl}

Let $\vert {\bf n} \rangle$, where ${\bf n}$ is any length-$N$ tuple of non-negative integers, be an orthonormal basis for our Hilbert space. Now suppose that the forms of $\hat{\bf z}$ and $\hat{\bf w}$ are
\begin{align}\label{eq:zw1}
\hat{z}_j &= \sum_{\bf n} \vert {\bf n} \rangle\langle {\bf n}+{\bf e}_j \vert, & \hat{w}_j &= \sum_{\bf n} (n_j+1) \vert {\bf n}+{\bf e}_j \rangle\langle {\bf n} \vert,
\end{align}
where 
\begin{equation}\label{eq:ej}
{\bf e}_j := (0, ..., 0, \underbrace{\;\quad 1 \quad\;}_{\text{position } j}, 0, ..., 0).
\end{equation}
It is straightforward to check that these operators satisfy both parts of Eq.~(\ref{eq:req}) with $\vert {\bf 0} \rangle$ being the ${\bf n} = {\bf 0}$ basis state. Additionally, the $\hat{\bf w}$ operators satisfy
\begin{equation}
[\hat{w}_j, \hat{w}_k] = 0,
\end{equation}
which can be used to simplify some expressions.

To concisely express the $\vert \psi(t) \rangle$ states, we first write $\vert {\bf n} \rangle$ as
\begin{equation}\label{eq:decomp}
\vert {\bf n} \rangle = \bigotimes_j \vert n_j \rangle.
\end{equation}
For instance, with $N=2$ we would have $\vert {\bf n} \rangle = \vert n_1 \rangle \otimes \vert n_2 \rangle$. Now,
\begin{align}\label{eq:cars}
\begin{split}
\vert \psi(t) \rangle = e^{{\bf z}(t) \cdot \hat{\bf w}} \vert {\bf 0} \rangle &= \bigotimes_j e^{z_j(t) \hat{w}_j} \vert 0 \rangle \\&= \bigotimes_j \sum_{n_j=0}^\infty [z_j(t)]^{n_j} \vert n_j \rangle.
\end{split}
\end{align}

Additional insight can be obtained after defining a number operator $\hat{n}$ by
\begin{equation}
\hat{n} \vert {\bf n} \rangle = \sum_j n_j \vert {\bf n} \rangle.
\end{equation}
Using $n$ to denote the $\hat{n}$ eigenvalues, states can be broken up into components of different $n$. For example, the $n=1$ component of $\vert \psi(t) \rangle$ is
\begin{equation}\label{eq:comp1}
\vert \psi_1(t) \rangle = \sum_j z_j(t) \vert {\bf e}_j \rangle,
\end{equation}
and the $n=2$ component is
\begin{equation}\label{eq:comp2}
\vert \psi_2(t) \rangle = \sum_{j \leq k} z_j(t) z_k(t) \vert {\bf e}_j+{\bf e}_k \rangle.
\end{equation}
Another observation is that $\hat{\bf z}$ decreases $n$ by one, except the $n=0$ component, $\vert {\bf 0} \rangle$, which it annihilates; and $\hat{\bf w}$ increases $n$ by one. Those familiar with the technique of Carleman linearization \cite{Carleman1932, Kow1991} should see that this particular linear embedding is just a representation of Carleman linearization in a Hilbert space. The development of an efficient quantum algorithm for dissipative nonlinear dynamical systems using Carleman linearization has been recently reported\cite{Dissipative}.

Finally, we consider output quantities. Any variable can be obtained as an inner product:
\begin{equation}
z_j(t) = \langle {\bf e}_j \vert \psi(t) \rangle.
\end{equation}
Similarly, a linear combination of variables
\begin{equation}
c(t) = \sum_j b_j z_j(t)
\end{equation}
is equivalent to $\langle c \vert \psi(t) \rangle$ with
\begin{equation}\label{eq:cex}
\vert c \rangle = \sum_j b^*_j \vert {\bf e}_j \rangle.
\end{equation}
Polynomials of the variables can also be obtained by adding $n>1$ components to $\vert c \rangle$. For instance, with $\vert c \rangle = \vert {\bf e}_j + {\bf e}_k \rangle$, $\langle c \vert \psi(t) \rangle = z_j(t) z_k(t)$. More generally, for any polynomial of the variables there is a specific state $\vert c \rangle$ for which $\langle c \vert \psi(t) \rangle$ evaluates that polynomial.

\subsection{Coherent States Embedding}
Now suppose that $\hat{\bf z} = \hat{\bf a}$ and $\hat{\bf w} = \hat{\bf a}^\dagger$, where $\hat{\bf a}$ are standard bosonic lowering operators. This version of linear embedding has been extensively explored by Kowalski \cite{Kow1994}. In this case Eq.~(\ref{eq:req}) becomes
\begin{align}
\hat{\bf a} \vert {\bf 0} \rangle &= {\bf 0}, & [\hat{a}_j, \hat{a}^\dagger_k] &= \delta_{jk},
\end{align}
which amounts to the standard commutation relations and the statement that $\vert {\bf 0} \rangle$ is the ground state. Additionally, we can use $\vert {\bf n} \rangle$ to denote the occupation basis and express the number operator as
\begin{equation}
\hat{n} = \sum_j \hat{a}^\dagger_j \hat{a}_j.
\end{equation}
Now the eigenvalues $n$ of $\hat{n}$ represent numbers of bosonic particles.

In this version of linear embedding, the evolution operator is
\begin{equation}
\hat{M} = \hat{\bf a}^\dagger \cdot {\bf F}(\hat{\bf a}),
\end{equation}
and the states are
\begin{align}\label{eq:cohs}
\begin{split}
\vert \psi(t) \rangle = e^{{\bf z}(t) \cdot \hat{\bf a}^\dagger} \vert {\bf 0} \rangle &= \bigotimes_j e^{z_j(t) \hat{a}^\dagger_j} \vert 0 \rangle \\&= \bigotimes_j \sum_{n_j=0}^\infty \frac{[z_j(t)]^{n_j}}{\sqrt{n_j!}} \vert n_j \rangle.
\end{split}
\end{align}
These are just coherent states; Eq.~(\ref{eq:eigz}) becomes
\begin{equation}
\hat{\bf a} \vert \psi(t) \rangle = {\bf z}(t) \vert \psi(t) \rangle.
\end{equation}

As in Sec.~\ref{sec:carl}, output quantities can be obtained with inner products. The only difference is that, for quadratic and higher-degree outputs, the $1/\sqrt{n_j!}$ factor in Eq.~(\ref{eq:cohs}) can have an effect. For example, $\vert c \rangle = \vert 2{\bf e}_j \rangle$ yields $\langle c \vert \psi(t) \rangle = z^2_j(t)/\sqrt{2}$ rather than $z^2_j(t)$.

While the states $\vert \psi(t) \rangle$ evolve linearly, this linear evolution is generally non-unitary with both the Carleman and coherent states embeddings. One simple way to see this is to note that $\vert \psi(t) \rangle$ [Eq.~(\ref{eq:cars}) or Eq.~(\ref{eq:cohs})] changes in normalization when $\vert z_j \vert$ for one $j$ changes. Non-unitary evolution presents difficulty for efficient quantum computation, so a linear embedding that gives unitary evolution may be preferable.

\subsection{Position-space Embedding}\label{sec:psp}

This time, take $\hat{\bf z} = \hat{\bf x}$ and $\hat{\bf w} = -i\hat{\bf p}$, where $\hat{\bf x}$ and $\hat{\bf p}$ are dimensionless versions of canonical position and momentum operators, respectively. Then
\begin{equation}
[\hat{z}_j, \hat{w}_k] = -i [\hat{x}_j, \hat{p}_k] = \delta_{jk},
\end{equation}
and both parts of Eq.~(\ref{eq:req}) are met with $\vert {\bf 0} \rangle = \vert {\bf x} = {\bf 0} \rangle$, the ${\bf x} = {\bf 0}$ position eigenstate. This version of linear embedding was introduced by Koopman and von Neumann\cite{Koopman, Neumann1, Neumann2}, and it has been studied in various forms by different authors \cite{Varadarajan, Chirikov, Alanson1992, Kow1997}. Since the eigenvalues of $\hat{\bf x}$ are real, the variables must be real, and to signify this we switch to using ${\bf x}$ to denote the variables instead of ${\bf z}$. To maintain the reality of ${\bf x}$, ${\bf F}(\cdot)$ must be a real function as well. The $\vert \psi(t) \rangle$ states are position eigenstates, and they can be expressed as
\begin{equation}\label{eq:psix}
\vert \psi(t) \rangle = \vert {\bf x}(t) \rangle = e^{-i {\bf x}(t) \cdot \hat{\bf p}} \vert {\bf x} = {\bf 0} \rangle.
\end{equation}
Since the translation operator $e^{-i {\bf x}(t) \cdot \hat{\bf p}}$ is unitary, the $\vert \psi(t) \rangle$ states have the same normalization as $\vert {\bf x} = {\bf 0} \rangle$.

The evolution operator is
\begin{equation}
\hat{M} = -i\hat{\bf p} \cdot {\bf F}(\hat{\bf x}).
\end{equation}
To study this embedding in the occupation basis $\vert {\bf n} \rangle$, we introduce
\begin{equation}\label{eq:ahat}
\hat{\bf a} = \frac{1}{\sqrt{2}} \left(\hat{\bf x} + i \hat{\bf p}\right).
\end{equation}
It is easily checked that these $\hat{\bf a}$ obey
\begin{align}
[\hat{a}_j, \hat{a}_k] &= 0, & [\hat{a}_j, \hat{a}^\dagger_k] &= \delta_{jk},
\end{align}
and thus they are standard bosonic lowering operators. Using the expressions for $\hat{\bf a}$ and $\hat{\bf a}^\dagger$ in the occupation basis, the evolution operator
\begin{equation}\label{eq:mpos}
\hat{M} = \frac{\hat{\bf a}^\dagger - \hat{\bf a}}{\sqrt{2}} \cdot {\bf F}\left(\frac{\hat{\bf a}^\dagger + \hat{\bf a}}{\sqrt{2}}\right)
\end{equation}
can also be expressed in the occupation basis.

To express $\vert \psi(t) \rangle$ [Eq.~(\ref{eq:psix})] in the occupation basis we need to relate the position and occupation bases. However, since Eq.~(\ref{eq:ahat}) is equivalent to the relation used in the quantum harmonic oscillator (QHO) problem with $\hbar = m\omega = 1$, the required result is well known: the one-dimensional QHO eigenstates in the position basis are
\begin{equation}
\langle x \vert n \rangle = e^{-x^2/2} \frac{H_n(x)}{\pi^{1/4}\sqrt{2^n n!}},
\end{equation}
where $H_n(x)$ are the physicists' Hermite polynomials. The form of $\vert \psi(t) \rangle \propto \vert {\bf x}(t) \rangle$ follows directly:
\begin{equation}\label{eq:pos_form}
\vert \psi(t) \rangle = \bigotimes_j \sum_{n_j=0}^\infty e^{-x_j(t)^2/2} \frac{H_{n_j}[x_j(t)]}{\sqrt{2^{n_j} n_j!}} \vert n_j \rangle,
\end{equation}
where we have chosen to drop the unimportant factors of $\pi^{-1/4}$.

Output quantities work somewhat differently in this version of linear embedding. The components of $\vert \psi(t) \rangle$ in the $n \leq b$ subspace, where $n$ denotes the eigenvalues of the total number operator $\hat{n}$, are polynomials of the variables of degree $\leq b$ times a factor of $e^{-{\bf x} \cdot {\bf x} / 2}$. Therefore, if we write
\begin{equation}\label{eq:overlap}
\langle c \vert \psi(t) \rangle = \exp\left[-\frac12 {\bf x}(t) \cdot {\bf x}(t)\right] p[{\bf x}(t)]
\end{equation}
for $\vert c \rangle$ a state in the $n \leq b$ subspace, then $p({\bf x})$ is a polynomial of the variables with degree $\leq b$. For example, $\vert c \rangle = \vert {\bf e}_j \rangle$ yields $p[{\bf x}(t)] = \sqrt{2}\, x_j(t)$. A potentially simpler way to express an output quantity is as the expectation value of an observable: $\langle g \rangle := \langle \psi(t) \vert g(\hat{\bf x}) \vert \psi(t) \rangle$ for some analytic function $g({\bf x})$. However, only $\langle c \vert \psi(t) \rangle$ can be evaluated within the $n \leq b$ subspace, which is crucial to the analysis in Sec.~\ref{sec:trunc}. The result [Eq.~(\ref{eq:tilde})] indicating that a truncated system can approximate the output for a sufficiently weak nonlinearity applies to $\langle c \vert \psi(t) \rangle$ values, not any $\langle g \rangle$ value. For this reason we exclusively study the $\langle c \vert \psi(t) \rangle$ way of expressing outputs.

The $e^{-{\bf x} \cdot {\bf x} / 2}$ factor in Eq.~(\ref{eq:overlap}) can complicate the representation of desired outputs, which are frequently polynomials of the variables. However, this issue is avoided if ${\bf x} \cdot {\bf F}({\bf x}) = 0$ for all ${\bf x}$. This ensures that $d_t ({\bf x} \cdot {\bf x}) = 0$, so the $e^{-{\bf x} \cdot {\bf x} / 2}$ factor in Eq.~(\ref{eq:overlap}) is constant. Additionally, this has a consequence for the form of the evolution operator: for ${\bf x} \cdot {\bf F}({\bf x}) = 0$ to hold for all ${\bf x}$, it must vanish identically, meaning that all terms are canceled. Therefore,
\begin{equation}\label{eq:vanish}
\hat{\bf r} \cdot {\bf F}(\hat{\bf r}) = 0
\end{equation}
holds for $\hat{\bf r}$ any vector of commuting operators. Now, $\hat{M}$ [Eq.~(\ref{eq:mpos})] can be expanded to give a sum of terms, each being proportional to a product of raising and lowering operators. This includes the terms that occur in the expansion of
\begin{equation}\label{eq:pure}
\frac{\hat{\bf a}^\dagger}{\sqrt{2}} \cdot {\bf F}\left(\frac{\hat{\bf a}^\dagger}{\sqrt{2}}\right) - \frac{\hat{\bf a}}{\sqrt{2}} \cdot {\bf F}\left(\frac{\hat{\bf a}}{\sqrt{2}}\right),
\end{equation}
which are made purely of raising or lowering operators. However, Eq.~(\ref{eq:vanish}) can be applied to find that Eq.~(\ref{eq:pure}) evaluates to zero. All the remaining terms in the expansion of $\hat{M}$ [Eq.~(\ref{eq:mpos})] have at least one raising and lowering operator. Also, supposing that ${\bf F}({\bf x})$ is a polynomial in ${\bf x}$ of degree $g$, then each term has a maximum of $g+1$ operators. Therefore, $\hat{M}$ couples between occupation basis components for which $n$, the $\hat{n}$ eigenvalue, differs by at most $g-1$.

The assumption that ${\bf x} \cdot {\bf F}({\bf x}) = 0$ for all ${\bf x}$ is non-trivial, but it holds in some important scenarios. For instance, it holds when
\begin{equation}\label{eq:preserve}
F_j({\bf x}) = W({\bf x}) (x_{j+1} - x_{j-1})
\end{equation}
for any $W({\bf x})$, with the indices handled cyclically: if $j$ is the last, then $j+1$ is the first. Equation (\ref{eq:preserve}) is a relevant form since it can be obtained when a first-order derivative in a partial differential equation is represented using centered differences and periodic boundary conditions.

Moreover, Eq.~(\ref{eq:preserve}) can be generalized to a large degree. First, instead of the index $j$ running over the usual vector components, it can run over any subset of the components, in any order, so long as they are handled cyclically. Components of ${\bf F}({\bf x})$ that are not in the subset are taken to be zero. Second, ${\bf F}({\bf x})$ can be any linear combination of terms of the described form, and each term can have a different $W({\bf x})$. As shown in Appendix~\ref{sec:app1}, the electrostatic Vlasov equation [Eq.~(\ref{eq:esv})] is an example of a partial differential equation that can be discretized to have this general form, implying that ${\bf x} \cdot {\bf F}({\bf x}) = 0$, where the variables ${\bf x}$ are the values of the distribution function on a grid.

Hereafter we assume that, when the position-space embedding is applied, the system satisfies the ${\bf x} \cdot {\bf F}({\bf x}) = 0$ condition. Therefore, the factor of $e^{-{\bf x} \cdot {\bf x}/2}$ is a constant, and it is convenient to divide this out of the state. Then $\vert \psi(t) \rangle$ becomes
\begin{equation}\label{eq:pos_state2}
\vert \psi(t) \rangle = \bigotimes_j \sum_{n_j=0}^\infty \frac{H_{n_j}[x_j(t)]}{\sqrt{2^{n_j} n_j!}} \vert n_j \rangle,
\end{equation}
and the output quantities [Eq.~(\ref{eq:overlap})] are updated to
\begin{equation}\label{eq:overlap2}
\langle c \vert \psi(t) \rangle = p[{\bf x}(t)],
\end{equation}
where, as before, $p({\bf x})$ is a polynomial of the variables with a form determined by the state $\vert c \rangle$.

Evolution in the position-space embedding can also be described in terms of a ``Hamiltonian" $\hat{H} = i\hat{M}$, and $\hat{H}$ can be written as
\begin{equation}\label{eq:ham1}
\hat{H} = \frac12 \left[\hat{\bf p} \cdot {\bf F}(\hat{\bf x}) + {\bf F}(\hat{\bf x}) \cdot \hat{\bf p}\right] + \frac12 \sum_j [\hat{p}_j, F_j(\hat{\bf x})].
\end{equation}
It is well known and easily verified that, for operators $\hat{x}$ and $\hat{p}$ satisfying $[\hat{x}, \hat{p}] = i$,
\begin{equation}
[\hat{p}, f(\hat{x})] = -i \frac{\partial f(\hat{x})}{\partial \hat{x}}
\end{equation}
for $f(\hat{x})$ any analytic function of $\hat{x}$. Applying this, we find that
\begin{equation}\label{eq:ham2}
\hat{H} = \frac12 \left[\hat{\bf p} \cdot {\bf F}(\hat{\bf x}) + {\bf F}(\hat{\bf x}) \cdot \hat{\bf p}\right] - \frac{i}{2} \text{div}\,{\bf F}(\hat{\bf x}),
\end{equation}
where $\text{div}\,{\bf F}(\hat{\bf x})$ is defined by replacing ${\bf x}$ with $\hat{\bf x}$ in $\text{div}\,{\bf F}({\bf x}) := \sum_j \partial F_j({\bf x}) / \partial {x}_j$. For many systems, $\text{div}\,{\bf F}({\bf x})$ evaluates to zero. If it does not, it is still possible to eliminate that term by extending the original system. In particular, consider two systems, each identical to the original, including in their initial conditions. Whenever $F_j({\bf x})$ contains $x_j$, we replace that $x_j$ occurrence with the corresponding variable from the other system. This has no impact on the dynamics, yet it results in the elimination of $\text{div}\,{\bf F}({\bf x})$ for the doubled system. We prefer this strategy over the elimination of that term through a redefinition of the states (e.g., as done by Kowalski\cite{Kow1997}) since the latter method introduces a factor of
\begin{equation}\label{eq:outfact}
\exp\left(\frac12 \int_0^t \text{div}\,{\bf F}[{\bf x}(\tau)] d\tau\right),
\end{equation}
which must be applied to obtain an output of the $\langle c \vert \psi(t) \rangle$ form, yet Eq.~(\ref{eq:outfact}) cannot be evaluated in general without the full solution ${\bf x}(t)$ to the system. We assume that the described extension is applied when necessary to get a system for which $\text{div}\,{\bf F}({\bf x})$ always vanishes. Then
\begin{equation}\label{eq:Hsym}
\hat{H} = \frac12 \left[\hat{\bf p} \cdot {\bf F}(\hat{\bf x}) + {\bf F}(\hat{\bf x}) \cdot \hat{\bf p}\right].
\end{equation}
In the occupation basis, Eq.~(\ref{eq:Hsym}) becomes
\begin{equation}
\hat{H} = \frac{i}{2} \left[\frac{\hat{\bf a}^\dagger - \hat{\bf a}}{\sqrt{2}} \cdot {\bf F}\left(\frac{\hat{\bf a}^\dagger + \hat{\bf a}}{\sqrt{2}}\right) + {\bf F}\left(\frac{\hat{\bf a}^\dagger + \hat{\bf a}}{\sqrt{2}}\right) \cdot \frac{\hat{\bf a}^\dagger - \hat{\bf a}}{\sqrt{2}}\right].
\end{equation}
Note that applying the standard rules for Hermitian conjugation to $\hat{H}$ now yields $\hat{H}^\dagger = \hat{H}$. As a consequence, the embedded evolution will be unitary when restricted to any finite-dimensional subspace of the occupation basis.

\subsection{Continuum Limit}

Although a finite $N$ will ultimately be required to perform computations, it is straightforward to formulate linear embedding for partial differential equations as well, e.g., as done by Kowalski\cite{Kow1994}. Consider a dynamical system of the form
\begin{equation}
d_t f({\bf q}, t) = F[{\bf q}, f(t)],
\end{equation}
where $F$ can contain derivatives and integrals of the function $f(t)$ with respect to the coordinates ${\bf q}$. If we introduce operators $\hat{f}_{\bf q}$ and $\hat{h}_{\bf q}$ and a state $\vert 0 \rangle$ such that
\begin{align}
\hat{f}_{\bf q} \vert 0 \rangle &= 0, & [\hat{f}_{\bf q}, \hat{h}_{{\bf q}'}] &= \delta({\bf q} - {\bf q}'),
\end{align}
then the states
\begin{equation}
\vert \psi(t) \rangle = \exp\left(\int f({\bf q},t) \hat{h}_{\bf q} d{\bf q} \right) \vert 0 \rangle
\end{equation}
are found to satisfy
\begin{equation}
\hat{f}_{\bf q} \vert \psi(t) \rangle = f({\bf q}, t) \vert \psi(t) \rangle
\end{equation}
and
\begin{equation}
d_t \vert \psi(t) \rangle = \hat{M} \vert \psi(t) \rangle,
\end{equation}
where
\begin{equation}
\hat{M} = \int \hat{h}_{\bf q} F({\bf q}, \hat{f}_{\bf q}) d{\bf q}.
\end{equation}
Any of the specific versions of linear embedding can then be applied to partial differential equations. For example, coherent states embedding applied to the electrostatic Vlasov equation for electrons [Eq.~(\ref{eq:esv})] gives
\begin{equation}
\vert \psi(t) \rangle = \exp\left(\int f({\bf x}, {\bf v}, t) \hat{a}_{{\bf x}, {\bf v}}^\dagger d{\bf x} d{\bf v}\right) \vert 0 \rangle
\end{equation}
and
\begin{multline}
\hat{M} = \int \hat{a}_{{\bf x}, {\bf v}}^\dagger \bigg[-{\bf v} \cdot \bm{\nabla} \hat{a}_{{\bf x}, {\bf v}} \\- \frac{1}{4\pi}\left(\int \hat{a}_{{\bf x}', {\bf v}'} \frac{{\bf x}-{\bf x}'}{\vert {\bf x}-{\bf x}' \vert^3} d{\bf x}' d{\bf v}'\right) \cdot \frac{\partial \hat{a}_{{\bf x}, {\bf v}}}{\partial {\bf v}} \bigg] d{\bf x} d{\bf v}.
\end{multline}

\section{Truncation of the Space}\label{sec:trunc}

Quantum computers can perform some linear computations such as matrix inversion with costs only logarithmic in the system size \cite{Harrow2009}, but that does not allow for handling a system of infinite size. Therefore, we seek to approximate a desired output quantity using linear evolution within some finite-dimensional subspace of the linear embedding space. Note that this is necessary even with a finite number of variables $N$. The infinite dimensionality of the linear embedding space is tied to the variables being represented with infinite precision, which is required to exactly represent continuous evolution in time.

The way in which we choose to truncate the linear embedding space makes a significant difference. For example, one can imagine a truncation of the position-space embedding (Sec.~\ref{sec:psp}) that replaces the infinite-precision variables ${\bf x}$ with finite-precision variables and represents $\hat{\bf p}$ with a finite difference matrix in the position basis. In this case the dimensionality of the truncated space scales as $(1/\epsilon)^N$, where $\epsilon$ is the chosen precision. Therefore, the space complexity, i.e., the required number of qubits, scales as $N \log_2(1/\epsilon)$. However, we will consider an alternative form of truncation and show that its space complexity can be only logarithmic in $N$. This potentially allows for the approximation of outputs with quantum computational complexity scaling only logarithmically with $N$, although accuracy is only ensured if the nonlinearity is sufficiently weak [Eq.~(\ref{eq:tilde})].

In what follows we assume that ${\bf F}(\cdot)$ is a polynomial of degree $g$ in the variables for some small integer $g$ and that ${\bf F}({\bf 0}) = {\bf 0}$, i.e., ${\bf F}(\cdot)$ does not have a constant term. Also, we assume that the desired computational output is a polynomial of degree $b$ in the variables at the final time for some small integer $b$. If the position-space embedding is used, then we add the requirement that ${\bf x} \cdot {\bf F}({\bf x}) = 0$ for all ${\bf x}$, as discussed in Sec.~\ref{sec:psp}. With the stated assumptions, a couple properties are shared by Carleman embedding, coherent states embedding, and position-space embedding. First, the output can be expressed as
\begin{equation}\label{eq:output}
c(t) = \langle c \vert \psi(t) \rangle
\end{equation}
for some state $\vert c \rangle$ belonging to the $n \leq b$ subspace of the occupation basis. As usual, $n$ is used to denote the eigenvalues of the total number operator $\hat{n}$. Second, the evolution operator $\hat{M}$ couples between occupation basis components for which $n$ differs by at most $g-1$.

Next, we rewrite the original dynamical system [Eq.~(\ref{eq:orig})] as
\begin{equation}\label{eq:split}
d_t {\bf z}(t) = A {\bf z}(t) + \eta {\bf G}[{\bf z}(t)],
\end{equation}
where $A$ is a matrix, $\eta$ is a constant, and ${\bf G}({\bf z})$ is purely nonlinear in ${\bf z}$. In particular, ${\bf G}({\bf z})$ is a degree-$g$ polynomial of the variables. Now the evolution operator can be decomposed as
\begin{align}\label{eq:mparts}
\hat{M} &= \hat{M}_0 + \eta\hat{M}_1, & \hat{M}_0 &:= \hat{\bf w} \cdot A \hat{\bf z}, & \hat{M}_1 &:= \hat{\bf w} \cdot {\bf G}(\hat{\bf z}).
\end{align}
By itself, $\hat{M}_0$ generates the linearized evolution of the original dynamical system, i.e., the evolution in the limit of $\eta \to 0$. Since $g=1$ for linear evolution, the terms in $\hat{M}_0$ do not couple between different $n$. The need to consider $ n > b$ components comes from $\eta\hat{M}_1$, which contains terms that change $n$ by up to $g-1$.

Now, we truncate the space such that only the
\begin{equation}\label{eq:subspace}
n \leq m := b + s(g-1)
\end{equation}
subspace is retained for some integer $s \geq 0$. Let $\tilde{c}(t)$ denote the approximation to $c(t)$ that is obtained by performing the embedded evolution in the truncated space and evaluating $\langle c \vert \psi(t) \rangle$ after some fixed, finite time $t$. Then
\begin{equation}\label{eq:tilde}
c(t) - \tilde{c}(t) = \mathcal{O}(\eta^{s+1}) \quad \text{as } \eta \to 0.
\end{equation}
We derive Eq.~(\ref{eq:tilde}) in Appendix~\ref{sec:app2}. Treating $\eta$ as a dimensionless parameter characterizing the strength of the nonlinearity, we expect $c(t) \approx \tilde{c}(t)$ for small $s$ if $\eta$ is sufficiently small. Note that how small $\eta$ needs to be for the approximation to be accurate can depend on $t$. If the dynamical system and output polynomial are actually linear, then the $s=0$ truncation is exact, and $m=1$. More generally, if the nonlinearity is sufficiently weak, which depends on the specific computation, then the output can be approximated with $s$, and thus also $m$, being of order one.

To find the dimensionality of the $n \leq m$ subspace, consider $N+1$ bins into which $m$ particles are placed. The first $N$ bins correspond to the numbers $n_j$ of Eq.~(\ref{eq:decomp}) while the last bin holds any extra particles. The subspace dimensionality is the number of unique placements of the particles into the bins, which is given by the binomial coefficient
\begin{equation}
\binom{N+m}{m} := \frac{(N+m)!}{N!m!}.
\end{equation}
Noting that
\begin{equation}
\frac{(N+m)!}{N!m!} = \prod_{j=0}^{m-1} \left(\frac{N}{m-j} + 1\right) \leq (N + 1)^m,
\end{equation}
the number of qubits required to represent the $n \leq m$ subspace is
\begin{equation}
\log_2 \tbinom{N+m}{m} = \mathcal{O}[m \ln(N+1)].
\end{equation}
Therefore, the space complexity of representing the Eq.~(\ref{eq:subspace}) subspace is
\begin{equation}\label{eq:spcp}
\mathcal{O}\{[b + s(g-1)] \ln(N+1)\}.
\end{equation}

\section{Efficiency}\label{sec:eff}

Some analysis of the efficiency of a quantum algorithm based on linear embedding is possible without specifying all the details. Let $\hat{M}'$ and $\vert \psi(0)' \rangle$ denote $\hat{M}$ and $\vert \psi(0) \rangle$, respectively, after restriction to the truncated subspace. Then the computation can be expressed as
\begin{equation}\label{eq:op1}
\hat{V}_c^\dagger e^{\hat{M}'t} \hat{V}_{\psi'},
\end{equation}
where $\hat{V}_{\psi'}$ and $\hat{V}_c$ are unitary operations that prepare states proportional to $\vert \psi(0)' \rangle$ and $\vert c \rangle$, respectively, from the computational starting state. After application of the Eq.~(\ref{eq:op1}) operation, the component along the computational starting state will be proportional to the output quantity, the details of which are determined by the chosen $\vert c \rangle$ state. The technique of amplitude estimation can then be applied to estimate this component using $\mathcal{O}(1/\varepsilon)$ iterations of the Eq.~(\ref{eq:op1}) operation, where $\varepsilon$ is an absolute accuracy for the output value\cite{Brassard2002}. In particular, that will yield, up to a normalization factor, an estimate of $\vert \tilde{c}(t) \vert$. However, the full complex value can also be estimated using a simple algorithm extension \cite{Engel2019}.

The $\vert \psi(0)' \rangle$ state is a sum of components with particle counts up to $m$. To represent this using qubits, we can use $m$ registers, each with $\left \lceil{\log_2(N+1)}\right \rceil$ qubits. Here $N+1$ appears instead of $N$ so that there is one extra basis state which can be used to indicate the absence of a particle: for the $n < m$ components there are $m-n$ unused registers which should be set to the extra basis state. This representation achieves the Eq.~(\ref{eq:spcp}) scaling. The $\vert c \rangle$ state is a sum of components (some of which may vanish) with particle counts up to $b \leq m$ [Eq.~(\ref{eq:subspace})]. Therefore, $\vert c \rangle$ can be represented in the same way as $\vert \psi(0)' \rangle$.

Overall efficiency requires efficient preparation of states proportional to $\vert \psi(0)' \rangle$ and $\vert c \rangle$. In particular, we want these operations to cost $\text{poly}(\log N)$. We are less concerned with the scaling with $m$, since we do not expect to achieve overall efficiency when $m$ is large. One basic state preparation strategy is to start by making a superposition over an ancilla register of $\left \lceil{\log_2(m+1)}\right \rceil$ qubits with the values of this register representing particle counts. Then, for each particle count $n \leq m$, the corresponding component of $\vert \psi(0)' \rangle$ [e.g., Eq.~(\ref{eq:comp2}) with $t=0$] or $\vert c \rangle$ [e.g., Eq.~(\ref{eq:cex})] is prepared. Whether that can be done with $\text{poly}(\log N)$ complexity will depend on the details of the initial state and output quantity, but it is possible in many cases, including when the maximum $\vert b_j \vert^2$ is a factor of $\mathcal{O}(1)$ larger than the average $\vert b_j \vert^2$, where $b_j$ is the prepared amplitude for state component $j$, assuming that each $b_j$ can be computed efficiently using the index $j$ \cite{Grover2002, Soklakov2006}.

In a quantum algorithm, prepared states must be normalized. Consequently, the output obtained from Eq.~(\ref{eq:op1}) will be based on the normalized versions of $\vert \psi(0)' \rangle$ and $\vert c \rangle$. A factor $\chi \zeta$, where $\chi := \sqrt{\langle \psi(0)' \vert \psi(0)' \rangle}$ and $\zeta := \sqrt{\langle c \vert c \rangle}$, must then be applied to the output to obtain the actual result. Therefore, to obtain the result to within some error tolerance $\delta$, the absolute accuracy $\varepsilon$ of the output value must be
\begin{equation}
\varepsilon = \frac{\delta}{\chi \zeta},
\end{equation}
and to extract the result to within error $\delta$ using amplitude estimation, $\mathcal{O}(\chi \zeta / \delta)$ iterations of the Eq.~(\ref{eq:op1}) operation are required\cite{Brassard2002}.

With position-space embedding the truncated system evolution is unitary, so it can potentially be performed efficiently using a Hamiltonian simulation algorithm such as the algorithm by Low and Chuang \cite{Low2017}. However, analysis of initial state normalization reveals a complication. The even-degree Hermite polynomials do not vanish at the origin, and therefore the initial state normalization receives large contributions from low-degree components. For example, some of the Eq.~(\ref{eq:pos_state2}) components are
\begin{equation}
\sqrt{2} \sum_j \left[x_j \vert {\bf e}_j \rangle + \left( x_j - \frac12\right) \vert 2 {\bf e}_j \rangle\right],
\end{equation}
and those contribute
\begin{equation}\label{eq:contribute}
2 \sum_j \left[x^2_j + \left(x_j-\frac12\right)^2\right]
\end{equation}
to $\langle \psi \vert \psi \rangle$. Equation (\ref{eq:contribute}) is minimized by $x_j = 1/4$ for all $j$, and the value at the minimum is $N/4$. Consequently, an $m=2$ truncation has $\chi^2 > N/4$. More generally, the scaling of $\chi^2$ with $N$ is $1/\chi^2 = \mathcal{O}[N^{-\lfloor{m/2}\rfloor}]$. This is a problem for the overall efficiency of an algorithm based on position-space embedding since performing amplitude estimation results in a cost factor of $\mathcal{O}(\chi \zeta / \delta)$.

With coherent states embedding, the initial state has a normalization of
\begin{equation}\label{eq:psin}
\sqrt{\langle \psi(0) \vert \psi(0) \rangle} = e^{\frac12 \vert{\bf z}(0)\vert^2}.
\end{equation}
Truncation of the space will mean that only a finite number of the components are kept, resulting in a lower initial state normalization. Specifically,
\begin{equation}
\chi^2 = \langle \psi(0)' \vert \psi(0)' \rangle = \sum_{n=0}^m \frac{\left[\frac12 \vert{\bf z}(0)\vert^2\right]^n}{n!}.
\end{equation}
To prevent $\chi$ from growing as some power of $N$, we need $\vert{\bf z}(0)\vert$ to not scale as a power of $N$. We can achieve that through a rescaling of the variables. In particular, we can switch from ${\bf z}$ to
\begin{equation}\label{eq:cov}
{\bf z}' = \gamma {\bf z}
\end{equation}
with a constant
\begin{equation}
\gamma = \mathcal{O}\left(\frac{1}{\vert {\bf z}(0) \vert}\right)
\end{equation}
so that $\gamma\vert{\bf z}(0)\vert = \mathcal{O}(1)$. The transformed system is
\begin{equation}\label{eq:zp}
d_t {\bf z}'(t) = \gamma {\bf F}\left[\frac{{\bf z}'(t)}{\gamma}\right],
\end{equation}
and the truncated initial state normalization becomes
\begin{equation}
\chi \leq \exp{\left\{\frac12 \vert\gamma{\bf z}(0)\vert^2\right\}} = \mathcal{O}(1).
\end{equation}

With Carleman embedding, a similar situation occurs. In this case the initial state normalization before truncation is
\begin{equation}
\sqrt{\langle \psi(0) \vert \psi(0) \rangle} = \prod_j (1-\vert z_j \vert^2)^{-1/2}
\end{equation}
when $\vert z_j \vert < 1$ for all $j$ and infinite otherwise, although the truncated state normalization will always be finite. Again, we can rescale the variables [Eq.~(\ref{eq:cov})], this time with
\begin{equation}
\gamma = \frac{1}{2\vert {\bf z}(0) \vert},
\end{equation}
where the factor of 2 is somewhat arbitrary. Next, we can apply
\begin{equation}
(1-\alpha)^{-1/2} (1-\beta)^{-1/2} \leq (1-\alpha-\beta)^{-1/2},
\end{equation}
which holds for non-negative $\alpha$ and $\beta$ with $\alpha+\beta < 1$, to bound the initial state normalization as
\begin{equation}
\chi \leq \left(1 - \vert \gamma {\bf z}(0) \vert^2\right)^{-1/2} = 2/\sqrt{3}.
\end{equation}
Thus, $\chi = \mathcal{O}(1)$ is achieved.

Applying rescaling [Eq.~(\ref{eq:cov})] to Carleman embedding or coherent states embedding results in a factor of $\gamma^n$ being applied to components of $\vert \psi(0)' \rangle$ for each particle count $n \leq m$. At the same time, to maintain the original output quantity, which is assumed to be a degree $b$ polynomial of the variables, the $n$-particle component of $\vert c \rangle$ is rescaled by $\gamma^{-n}$ for each $n \leq b$. The general form of the output quantity is
\begin{equation}
c(t) = \sum_{n=0}^b \sum_{j_1,...,j_n} C^n_{j_1,...,j_n} \prod_{i=1}^n z_{j_i}(t),
\end{equation}
where $C^n$ is a rank-$n$ tensor. Suppose that, prior to any rescaling of the variables, the variable values are bounded as $N \to \infty$, while the entries of $C^n$ scale as $\mathcal{O}(N^{-n})$. For example, the variables can represent a bounded distribution function over a grid of size $N$, and the output quantity can approximate some integral over the distribution function by summing over the values on the grid. Then $\vert{\bf z}(0)\vert = \mathcal{O}(\sqrt{N})$, and a rescaling with $\gamma \propto 1/\sqrt{N}$ will result in $\chi = \mathcal{O}(1)$ and $C^n_{j_1,...,j_n} = \mathcal{O}(N^{-n/2})$. Now, the components of $\vert c \rangle$ in the occupation basis are proportional to the entries of the $C$ tensors (in a manner independent of $N$). Therefore, the scaling of $\zeta^2$ with $N$ can be obtained by summing all squared absolute entries of $C^n$ for each $n$, which yields $\zeta = \mathcal{O}(1)$ with respect to $N$ in this case.

So Carleman embedding and coherent states embedding can potentially avoid having the cost factor $\chi \zeta$ grow as a power of $N$, but with these embeddings the linear evolution given by $e^{\hat{M}'t}$ is generally not unitary. We can perform this non-unitary evolution using a QLSA, e.g., as done by Berry \textit{et al} \cite{ODE2017}. However, the more non-unitary the evolution is, the higher the costs will be. Of particular concern is that the condition number $\kappa$ of the evolution may grow exponentially with the simulation time $t$, forcing the costs to also grow exponentially with $t$.

But there is reason to suspect that, in some cases, an $\exp(t)$ cost scaling can be avoided. With coherent states embedding, if the system is given by Eq.~(\ref{eq:split}) with the matrix $A$ being anti-Hermitian, then $\hat{M}_0$ [Eq.~(\ref{eq:mparts})] is formally anti-Hermitian:
\begin{equation}
\hat{M}_0^\dagger = \hat{\bf a}^\dagger \cdot A^\dagger \hat{\bf a} = -\hat{\bf a}^\dagger \cdot A \hat{\bf a} = -\hat{M}_0.
\end{equation}
Additionally, if we order the basis states based on their particle count $n$, from low to high, then $\hat{M}$ takes on a block-upper-triangular structure, with one block for each $n$. This occurs because $\hat{M}_1$ always decreases $n$, while $\hat{M}_0$ leaves $n$ constant. All the blocks along the diagonal are anti-Hermitian since they are components of $\hat{M}_0$. Then a block-diagonal, unitary transformation can be applied to diagonalize $\hat{M}_0$ while maintaining the overall block-upper-triangular structure. That will result in an upper-triangular matrix with purely imaginary entries along the diagonal. The same holds for the matrix $\hat{M}'$ obtained by truncating $\hat{M}$ to the $n \leq m$ subspace, which implies that the eigenvalues of $\hat{M}'$ are purely imaginary. As a consequence, the condition number $\kappa$ of the evolution $e^{\hat{M}'t}$ will remain bounded as $t \to \infty$ if $\hat{M}'$ is diagonalizable. If $\hat{M}'$ happens to be non-diagonalizable, the block structure of $\hat{M}'$ limits the size of any Jordan chain to $m$, which leads to a bound of $\kappa = \mathcal{O}(t^{2(m-1)})$. A worse scaling with $t$ could still occur if the truncation degree $m$ needed to obtain accurate results increases with $t$, which will depend on the details of the classical dynamical system.

Another important question is how the costs of implementing the $e^{\hat{M}'t}$ evolution scale with $N$. With the same assumptions as in Sec.~\ref{sec:trunc}, the general form of the classical system is
\begin{equation}\label{eq:expsys}
d_t z_{j_0}(t) = \sum_{r=1}^g \sum_{j_1,...,j_r} A^r_{j_0,...,j_r} \prod_{i=1}^r z_{j_i}(t),
\end{equation}
where $A^r$ is a rank-$(r+1)$ tensor. Without any loss of generality, we take $A^r$ to be symmetric in its last $r$ indices. Now suppose that all the $A^r$ are $q$-sparse in their first index. In other words, for fixed $j_1,...,j_r$, the number of $j_0$ indices such that $A^r_{j_0,...,j_r} \neq 0$ is at most $q$. For many systems, $q$ will be small, including the electrostatic Vlasov system [Eq.~(\ref{eq:esv})] discretized on a grid in ${\bf x}$ and ${\bf v}$. The evolution operator $\hat{M}'$ associated with Eq.~(\ref{eq:expsys}) is not generally sparse when expressed in the occupation basis, even for small $q$, but it does have a structure that may allow for approximating the $e^{\hat{M}'t}$ evolution efficiently; this is explored further in Appendix~\ref{sec:app3}. Still, the costs grow with the size of the entries of $\hat{M}'$, and thus with the entries of $A^r$. In particular, there is a cost factor (explained in Appendix~\ref{sec:app3}) of
\begin{equation}\label{eq:nmf}
\sqrt{\max_{j_0} \sum_r \sum_{j_1,...,j_r} \vert \gamma^{1-r} A^r_{j_0,...,j_r} \vert^2},
\end{equation}
where a rescaling of the form described earlier [Eqs.(\ref{eq:cov}) and (\ref{eq:zp})] is assumed to have been applied. Whether Eq.~(\ref{eq:nmf}) scales as a power of $N$ still depends on the classical system, and on the details of how the system is being varied as $N$ is increased. For systems obtained by discretizing a local, nonlinear partial differential equation, this scaling may be improved by switching to Fourier space (within which the system is no longer local).

\section{Discussion}\label{sec:dis}

Numerical simulations of nonlinear dynamical systems on regular computers have costs at least linear in the number of variables $N$. Kinetic plasma simulation is especially expensive since very many variables (e.g., phase space grid points, Fourier modes, or basis functions) are needed to accurately represent the six-dimensional particle distribution function. This motivates investigating whether a quantum computer might be able to perform the same computation with costs sublinear in $N$. In particular, we have focused on computations that may be possible with costs growing only logarithmically with $N$.

We have investigated the following approach. The nonlinear dynamical system is first mapped to an infinite-dimensional linear dynamical system using linear embedding. The linear system is then expressed in the occupation basis and truncated to a system of size $\mathcal{O}[(N+1)^m]$ for a small integer $m$. If a quantum algorithm can obtain an exponential speedup for the resulting linear computation, then the costs will scale with $N$ as $\mathcal{O}[m \ln(N+1)]$ on a quantum computer. However, whether this logarithmic scaling with $N$ is achievable will depend on the classical dynamical system being studied.

There is no guarantee that a truncated linear system will give a good approximation to the exact nonlinear evolution in general. We noted that the truncation will give correct results up to an order $s \propto m-b$ in a parameter $\eta$ associated with the strength of the nonlinearity. This implies that a truncation will be accurate if $\eta$ is sufficiently small. However, whether that is the case is a difficult question that depends on the details of the problem being studied. Similarly, when $\eta$ is larger, at what order the terms become negligible, if ever, will be problem dependent. Whether a low-$m$ truncation can approximate the output of any difficult, nonlinear kinetic plasma problem is also an open question.

Additionally, it could be that the non-unitarity of the evolution of the truncated linear system will make the quantum computation inefficient. In particular, if the costs end up being logarithmic in $N$ but exponential in the simulation time $t$, that will typically not give a speedup over the $\mathcal{O}(N t)$ costs for doing the simulation classically. We conjecture that there are certain classes of nonlinear systems for which an $\exp(t)$ scaling can be avoided. The structure and conservation properties of physical systems such as the Vlasov-Poisson system might be relevant here, and this is an interesting topic that we leave for future work.

\begin{acknowledgments}
Research was supported in part by the U.S. Department of Energy under Grant No. DE-SC0020393.
\end{acknowledgments}

\appendix

\section{}\label{sec:app1}

Here we show that a very simple discretization of the electrostatic Vlasov equation [Eq.~(\ref{eq:esv})] satisfies the ${\bf x} \cdot {\bf F}({\bf x}) = 0$ condition discussed in Sec.~\ref{sec:psp}. Let the variable $f_{{\bf j}, {\bf k}}$ represent the value of the distribution function at ${\bf x} = {\bf j} \Delta x + {\bf x}_0$ and ${\bf v} = {\bf k} \Delta v + {\bf v}_0$ for some choices of $\Delta x$, $\Delta v$, ${\bf x}_0$, and ${\bf v}_0$. The components of ${\bf j}$ and ${\bf k}$ are integers between zero and some maximum, which can be different for each component. Also, for each expression indexed by ${\bf j}$ and ${\bf k}$, there is a corresponding vector representation obtained by collapsing the ${\bf j}$ and ${\bf k}$ indices into a single index in some fixed manner. Then we seek to show that ${\bf f} \cdot {\bf F}({\bf f}) = 0$, which we can do by breaking ${\bf F}({\bf f})$ up into a number of terms, ${\bf F}({\bf f}) = \sum_i {\bf F}_i({\bf f})$, and showing that ${\bf f} \cdot {\bf F}_i({\bf f}) = 0$ for each $i$.

Using centered differences, the $-v_i \nabla_i f({\bf x}, {\bf v})$ term takes the form
\begin{equation}
-({\bf k} \Delta v + {\bf v}_0)_i \frac{f_{{\bf j}+{\bf e}_i, {\bf k}}-f_{{\bf j}-{\bf e}_i, {\bf k}}}{2 \Delta x},
\end{equation}
with ${\bf e}_i$ defined as in Eq.~(\ref{eq:ej}). We assume periodic boundary conditions so that the indices are handled cyclically, i.e., $j_i + 1$ and $j_i - 1$ wrap around when they fall outside the range of valid indices. The evaluation of ${\bf f} \cdot {\bf F}_i({\bf f})$ for this term gives
\begin{multline}
-\sum_{{\bf j}, {\bf k}} ({\bf k} \Delta v + {\bf v}_0)_i f_{{\bf j}, {\bf k}} \frac{f_{{\bf j}+{\bf e}_i, {\bf k}}-f_{{\bf j}-{\bf e}_i, {\bf k}}}{2 \Delta x}\\
=\sum_{{\bf j}, {\bf k}} \frac{({\bf k} \Delta v + {\bf v}_0)_i}{2\Delta x} (f_{{\bf j}-{\bf e}_i, {\bf k}} f_{{\bf j}, {\bf k}}-f_{{\bf j}, {\bf k}} f_{{\bf j}+{\bf e}_i, {\bf k}}) = 0,
\end{multline}
where the last equality holds because each $f_{{\bf j}, {\bf k}} f_{{\bf j}+{\bf e}_i, {\bf k}}$ appears twice in the sum with opposite signs and the same prefactor. The key to this cancellation, besides the cyclic handling of the indices, is that the prefactor $({\bf k} \Delta v + {\bf v}_0)_i/(2\Delta x)$ does not depend on the position index $j_i$, which is the only index that differs between $f_{{\bf j}-{\bf e}_i, {\bf k}} f_{{\bf j}, {\bf k}}$ and $f_{{\bf j}, {\bf k}} f_{{\bf j}+{\bf e}_i, {\bf k}}$.

The same sort of cancellation occurs for the nonlinear terms of Eq.~(\ref{eq:esv}), which take the form
\begin{equation}\label{eq:tnon}
- \frac{1}{4\pi}\left(\sum_{{\bf j}', {\bf k}'} f_{{\bf j}', {\bf k}'} \frac{({\bf j}-{\bf j}')_i}{\vert {\bf j}-{\bf j}' \vert^3} \frac{(\Delta x \Delta v)^3}{(\Delta x)^2}\right) \frac{f_{{\bf j}, {\bf k}+{\bf e}_i}-f_{{\bf j}, {\bf k}-{\bf e}_i}}{2 \Delta x}.
\end{equation}
This time the prefactor is a function of the variables and is proportional to the electric field component $E_i({\bf x})$. We can express ${\bf f} \cdot {\bf F}_i({\bf f})$ for the Eq.~(\ref{eq:tnon}) term as
\begin{equation}
\sum_{{\bf j}, {\bf k}}\frac{E_i({\bf j} \Delta x + {\bf x}_0)}{2\Delta v} (f_{{\bf j}, {\bf k}-{\bf e}_i} f_{{\bf j}, {\bf k}}-f_{{\bf j}, {\bf k}} f_{{\bf j}, {\bf k}+{\bf e}_i}) = 0,
\end{equation}
where all terms cancel because the electric field does not depend on the velocity index $k_i$. Therefore, ${\bf f} \cdot {\bf F}_i({\bf f}) = 0$ holds for all the terms in our discretized form of Eq.~(\ref{eq:esv}).

\section{}\label{sec:app2}

We now derive Eq.~(\ref{eq:tilde}), using the definitions and assumptions from Sec.~\ref{sec:trunc}. Let $\hat{M}'_0$ and $\hat{M}'_1$ denote the operators obtained by restricting $\hat{M}_0$ and $\hat{M}_1$, respectively, to the $n \leq m$ subspace. Since $\hat{M}'_0$ and $\hat{M}'_1$ are finite-dimensional operators with finite entries, they have finite spectral norms, which we denote as $\Lambda_0$ and $\Lambda_1$, respectively. The output approximation is
\begin{equation}\label{eq:sert}
\tilde{c}(t, \eta) = \sum_{j=0}^\infty \frac{\langle c \vert [(\hat{M}'_0 + \eta \hat{M}'_1)t]^j \vert \psi'(0) \rangle}{j!},
\end{equation}
where $\vert \psi'(0) \rangle$ is the initial state restricted to the $n \leq m$ subspace. $\tilde{c}_j(t, \eta)$ is the same as $\tilde{c}(t)$ from Sec.~\ref{sec:trunc}, now just with the $\eta$ dependence made explicit. Alternatively, we can write $\tilde{c}(t, \eta)$ as a power series in $\eta$:
\begin{equation}\label{eq:sern}
\tilde{c}(t, \eta) = \sum_{j=0}^\infty \tilde{c}_j(t) \eta^j.
\end{equation}
If we collect the Eq.~(\ref{eq:sert}) terms that have $j$ factors of $\eta$ and apply the subadditive and submultiplicative properties of spectral norms, we obtain a bound
\begin{equation}
\begin{split}
\vert \tilde{c}_j(t) \eta^j \vert &\leq \vert\eta \Lambda_1 t\vert^j \sum_{k=0}^\infty \vert \Lambda_0 t \vert^k \binom{j+k}{j} \frac{1}{(j+k)!}\\
&= \frac{\vert \eta \Lambda_1 t \vert^j}{j!} \exp(\vert \Lambda_0 t \vert),
\end{split}
\end{equation}
from which it follows that the Eq.~(\ref{eq:sern}) series is convergent for all finite $t$ and $\eta$. Therefore,
\begin{equation}\label{eq:app_ser}
\tilde{c}(t, \eta) = \sum_{j=0}^s \tilde{c}_j(t) \eta^j + \mathcal{O}(\eta^{s+1}) \quad \text{as } \eta \to 0
\end{equation}
holds for finite $t$ and any integer $s \geq 0$. Meanwhile, the exact output quantity $c(t, \eta)$ [equal to $c(t)$ from Sec.~\ref{sec:trunc}] can be expressed as the right side of Eq.~(\ref{eq:sert}) without the primes or as
\begin{equation}\label{eq:asym}
c(t, \eta) = \sum_{j=0}^\infty c_j(t) \eta^j,
\end{equation}
but in this case the prior argument for convergence does not apply since $\hat{M}_0$ and $\hat{M}_1$ do not have finite spectral norms. Instead, we apply a result from the theory of first-order, ordinary differential equations: for an initial value problem given by
\begin{align}
d_t {\bf z}(t, \eta) &= {\bf F}({\bf z}, \eta), & {\bf z}(0, \eta) &= {\bf w},
\end{align}
where ${\bf F}({\bf z}, \eta)$ is infinitely differentiable with respect to $\eta$, if there is a unique $\eta=0$ solution ${\bf z}_0(t)$ for $t \in [0, T]$ with $T$ finite, then for any integer $s \geq 0$, there exists functions ${\bf z}_j(t)$ such that
\begin{equation}\label{eq:ptex}
{\bf z}(t, \eta) = \sum_{j=0}^s {\bf z}_j(t) \eta^j + \mathcal{O}(\eta^{s+1}) \quad \text{as } \eta \to 0
\end{equation}
holds for $t \in [0, T]$; this is a particular case of what Hoppensteadt\cite{Perturbations} calls the regular perturbation theorem. The need to consider a finite time interval arises because the series in $\eta$ might not converge in the limit of $t \to \infty$. These conditions are met: ${\bf F}({\bf z}, \eta) := A {\bf z} + \eta {\bf G}({\bf z})$ is infinitely differentiable in $\eta$, the $\eta=0$ solution is just ${\bf z}_0(t) = e^{At} {\bf w}$, and we only consider finite simulation times.

Next, since $c(t, \eta)$ is assumed to be a specified polynomial of ${\bf z}(t, \eta)$, we can plug Eq.~(\ref{eq:ptex}) into this polynomial to obtain
\begin{equation}\label{eq:ex_ser}
c(t, \eta) = \sum_{j=0}^s c_j(t) \eta^j + \mathcal{O}(\eta^{s+1}) \quad \text{as } \eta \to 0,
\end{equation}
where the functions $c_j(t)$ must be the same as in Eq.~(\ref{eq:asym}) due to the uniqueness of asymptotic expansions. Finally, we use that $\tilde{c}_j(t) = c_j(t)$ for $j \leq s$ with the value of $s$ from Eq.~(\ref{eq:subspace}). This is true because it takes $s+1$ applications of $\hat{M}_1$ to couple from $\vert c \rangle$ to any component with $n > m$. Therefore, all terms in
\begin{equation}
c(t, \eta) = \langle c \vert \exp(\hat{M}_0 + \eta \hat{M}_1) \vert \psi(0) \rangle
\end{equation}
that are affected by the truncation [Eq.~(\ref{eq:subspace})] have at least $s+1$ factors of $\eta$, which implies that $\tilde{c}_j(t) = c_j(t)$ for $j \leq s$. Combining this result with Eqs. (\ref{eq:app_ser}) and (\ref{eq:ex_ser}) yields Eq.~(\ref{eq:tilde}). Of course, in practice we need the approximation to be accurate for particular values of $\eta$ and $t$, while Eq.~(\ref{eq:tilde}) only ensures accuracy for sufficiently small $\eta$, in a manner that can depend on $t$.

\section{}\label{sec:app3}

Here we outline a strategy for implementing $e^{\hat{M}'t}$ for classical systems of the form given in Eq.~(\ref{eq:expsys}) with each tensor being $q$-sparse in its first index. Let $M'$ denote a matrix representing $\hat{M}'$ in the occupation basis. We consider only Carleman embedding and coherent states embedding. In both cases,
\begin{align}
\hat{w}_j \vert {\bf n} \rangle &\propto \vert {\bf n}+{\bf e}_j \rangle, &\hat{z}_j \vert {\bf n} \rangle &\propto \vert {\bf n}-{\bf e}_j \rangle,
\end{align}
and $\hat{z}_j$ annihilates $\vert {\bf n} \rangle$ when $n_j = 0$. Additionally, the $\hat{\bf z}$ operators occur on the right in $\hat{M}$, and every term has at least one $\hat{\bf z}$ factor:
\begin{equation}
\hat{M} = \sum_{r=1}^g \sum_{j_0,...,j_r} \hat{w}_{j_0}A^r_{j_0,...,j_r} \prod_{i=1}^r \hat{z}_{j_i}.
\end{equation}
This allows us to bound the number of different occupation basis components in $\hat{M} \vert {\bf n} \rangle$, where $\vert {\bf n} \rangle$ lies within the truncated subspace, meaning that $\sum_j n_j \leq m$. The number of ways that a number of particles between 1 and $g$ can be removed from $\vert {\bf n} \rangle$ is bounded by
\begin{equation}\label{eq:nrem}
\sum_{r=1}^g \binom{m}{r} \leq \sum_{r=1}^g m^r \leq (m+1)^g.
\end{equation}
For each removal of particles counted in Eq.~(\ref{eq:nrem}), there is a corresponding product of $\hat{\bf z}$ operators which can appear in the terms of $\hat{M}$ with up to $q$ different $\hat{\bf w}$ factors. Therefore, the number of  occupation basis components in $\hat{M} \vert {\bf n} \rangle$, which equals the number of non-zero entries in a column of $M'$, is bounded by
\begin{equation}\label{eq:sparse}
q (m+1)^g.
\end{equation}
Since $N$ does not appear in Eq.~(\ref{eq:sparse}), we consider $M'$ to be sparse along its columns. This is not generally true along its rows, i.e., there can be $\text{poly}(N)$ non-zero entries in a row of $M'$.

Evolution by $M'$ can be converted into a matrix inversion problem using the technique of Berry \textit{et al} \cite{ODE2017}. The matrix $L$ to be inverted has up to two more non-zero entries in each column than the evolution matrix.\cite{ODE2017} To perform the matrix inversion using a QLSA, the Hermitian matrix
\begin{equation}\label{eq:invert}
\tilde{L} = \begin{pmatrix}
0 & L\\
L^\dagger & 0
\end{pmatrix}
\end{equation}
is inverted instead \cite{Harrow2009}. While $L$ is sparse along its columns, $\tilde{L}$ is not sparse. However, if Hamiltonian simulation of $\tilde{L}$ can be performed efficiently, then that can be used to implement an efficient QLSA\cite{Harrow2009,QLSA2017}.

The Hamiltonian simulation algorithm by Low and Chuang \cite{Qubitization} can be efficient for some non-sparse Hamiltonians. This includes when the Hamiltonian $H$ can be expressed as
\begin{equation}\label{eq:encoding}
H_{jk} = \Gamma \langle \chi_{j} \vert \varphi_{k} \rangle,
\end{equation}
where the states $\vert \chi_{j} \rangle$ and $\vert \varphi_{k} \rangle$ can be efficiently prepared, and $\Gamma$ is not too large: the costs grow linearly with $\Gamma$ \cite{Engel2019, LowSpectral}. We now examine how that can be applied to the simulation of $\tilde{L}$.

Let $W$ denote the size of $L$ and $d \leq q (m+1)^g+2$ the maximum number of non-zero entries in any column of $L$. We introduce a unitary operation $\hat{O}$ such that
\begin{equation}\label{eq:hato}
\hat{O} \vert y \rangle \vert l \rangle \vert k \rangle \vert x \rangle := \vert y \rangle \vert l \rangle \vert k \rangle \vert x \oplus j \rangle,
\end{equation}
where $j$ is the row in $L$ with the $l$\textsuperscript{th} occurrence of a non-zero entry along column $k$; if there are not $l$ occurrences or $k \geq W$, then $j$ can be arbitrary. In Eq.~(\ref{eq:hato}) and below, we take the registers to have dimensions of 2, $d$, $2W$, and $W$, from left to right. Next, we introduce a unitary operation $\hat{P}$ that acts as
\begin{equation}
\hat{P}\vert 0 \rangle \vert 0 \rangle \vert 0 \rangle \vert j \rangle = \frac{1}{\Lambda} \vert 0 \rangle \sum_i L^*_{j k_i} \vert l_i \rangle \vert k_i \rangle \vert j \rangle + \vert 1 \rangle \vert \phi_j \rangle
\end{equation}
for some constant $\Lambda$, where $L_{jk_i}$ is the $l_i$\textsuperscript{th} non-zero entry along column $k_i$ of $L$, and $\vert \phi_j \rangle$ is any superposition of $\vert l \rangle \vert k \rangle \vert x \rangle$ components with $k < W$. The sets $\{l_i\}$ and $\{k_i\}$ depend implicitly on $j$; in particular, $\{k_i\} = \{k | L_{jk} \neq 0\}$. Then
\begin{equation}
\hat{O}\hat{P} \vert 0 \rangle \vert 0 \rangle \vert 0 \rangle \vert j \rangle = \frac{1}{\Lambda} \vert 0 \rangle \sum_i L^*_{jk_i} \vert l_i \rangle \vert k_i \rangle \vert 0 \rangle + \vert 1 \rangle \vert \phi'_j \rangle,
\end{equation}
where the $\vert \phi'_j \rangle$ states satisfy $\hat{O} \vert 1 \rangle \vert \phi_j \rangle = \vert 1 \rangle \vert \phi'_j \rangle$.

Now we introduce states
\begin{equation}
\vert \chi_j \rangle =
\begin{cases}
\hat{O}\hat{P} \vert 0 \rangle \vert 0 \rangle \vert 0 \rangle \vert j \rangle & \mbox{for } j < W, \\
\frac{1}{\sqrt{d}} \sum_{l=0}^{d-1} \vert 0 \rangle \vert l \rangle \vert j\rangle \vert 0 \rangle & \mbox{ otherwise.}
\end{cases}
\end{equation}
Also, we define $\vert \varphi_j \rangle$ as the state obtained by applying the operation $+ W \,(\text{mod } 2W)$ to the third register of $\vert \chi_j \rangle$. Then $\vert \chi_j \rangle$ and $\vert \varphi_k \rangle$ satisfy Eq.~(\ref{eq:encoding}) with $H = \tilde{L}$ and $\Gamma = \sqrt{d} \Lambda$. Moreover, they are straightforward to prepare using an implementation of the $\hat{O}\hat{P}$ operation.

The $\hat{P}$ operation requires preparing a superposition containing amplitudes that are proportional to the entries of the classical system tensors $A^r$. Consequently, the cost to implement the $\hat{P}$ operation will depend on details of the classical system. However, there is one result that holds generally: the unitarity of $\hat{P}$ gives a lower bound on the constant $\Lambda$. It cannot be less than the maximum normalization of any row in $L$. This is the source of the Eq.~(\ref{eq:nmf}) cost factor. Factors that grow as powers of $m$ and depend on whether Carleman embedding or coherent state embedding is applied have been dropped from Eq.~(\ref{eq:nmf}) for simplicity.

\section*{Data Availability}
Data sharing is not applicable to this article as no new data were created or analyzed in this study.

\section*{References}
%


\begin{thebibliography}{35}%
\makeatletter
\providecommand \@ifxundefined [1]{%
 \@ifx{#1\undefined}
}%
\providecommand \@ifnum [1]{%
 \ifnum #1\expandafter \@firstoftwo
 \else \expandafter \@secondoftwo
 \fi
}%
\providecommand \@ifx [1]{%
 \ifx #1\expandafter \@firstoftwo
 \else \expandafter \@secondoftwo
 \fi
}%
\providecommand \natexlab [1]{#1}%
\providecommand \enquote  [1]{``#1''}%
\providecommand \bibnamefont  [1]{#1}%
\providecommand \bibfnamefont [1]{#1}%
\providecommand \citenamefont [1]{#1}%
\providecommand \href@noop [0]{\@secondoftwo}%
\providecommand \href [0]{\begingroup \@sanitize@url \@href}%
\providecommand \@href[1]{\@@startlink{#1}\@@href}%
\providecommand \@@href[1]{\endgroup#1\@@endlink}%
\providecommand \@sanitize@url [0]{\catcode `\\12\catcode `\$12\catcode
  `\&12\catcode `\#12\catcode `\^12\catcode `\_12\catcode `\%12\relax}%
\providecommand \@@startlink[1]{}%
\providecommand \@@endlink[0]{}%
\providecommand \url  [0]{\begingroup\@sanitize@url \@url }%
\providecommand \@url [1]{\endgroup\@href {#1}{\urlprefix }}%
\providecommand \urlprefix  [0]{URL }%
\providecommand \Eprint [0]{\href }%
\providecommand \doibase [0]{https://doi.org/}%
\providecommand \selectlanguage [0]{\@gobble}%
\providecommand \bibinfo  [0]{\@secondoftwo}%
\providecommand \bibfield  [0]{\@secondoftwo}%
\providecommand \translation [1]{[#1]}%
\providecommand \BibitemOpen [0]{}%
\providecommand \bibitemStop [0]{}%
\providecommand \bibitemNoStop [0]{.\EOS\space}%
\providecommand \EOS [0]{\spacefactor3000\relax}%
\providecommand \BibitemShut  [1]{\csname bibitem#1\endcsname}%
\let\auto@bib@innerbib\@empty
\bibitem [{\citenamefont {Murty}(1983)}]{Murty1983}%
  \BibitemOpen
  \bibfield  {author} {\bibinfo {author} {\bibfnamefont {S.}~\bibnamefont
  {Murty}},\ }\bibfield  {title} {\enquote {\bibinfo {title} {Engineering
  computations at the {N}ational {M}agnetic {F}usion {E}nergy {C}omputer
  {C}enter},}\ }\href {https://doi.org/10.13182/FST83-A22771} {\bibfield
  {journal} {\bibinfo  {journal} {Nuclear Technology/Fusion}\ }\textbf
  {\bibinfo {volume} {4}},\ \bibinfo {pages} {25--32} (\bibinfo {year}
  {1983})}\BibitemShut {NoStop}%
\bibitem [{\citenamefont {Service}(2018)}]{Service2018}%
  \BibitemOpen
  \bibfield  {author} {\bibinfo {author} {\bibfnamefont {R.}~\bibnamefont
  {Service}},\ }\bibfield  {title} {\enquote {\bibinfo {title} {Design for
  {U}.{S}. exascale computer takes shape},}\ }\href
  {https://doi.org/10.1126/science.359.6376.617} {\bibfield  {journal}
  {\bibinfo  {journal} {Science}\ }\textbf {\bibinfo {volume} {359}},\ \bibinfo
  {pages} {617} (\bibinfo {year} {2018})}\BibitemShut {NoStop}%
\bibitem [{\citenamefont {Preskill}(2018)}]{Preskill2018}%
  \BibitemOpen
  \bibfield  {author} {\bibinfo {author} {\bibfnamefont {J.}~\bibnamefont
  {Preskill}},\ }\bibfield  {title} {\enquote {\bibinfo {title} {Quantum
  computing in the {NISQ} era and beyond},}\ }\href
  {https://doi.org/10.22331/q-2018-08-06-79} {\bibfield  {journal} {\bibinfo
  {journal} {{Quantum}}\ }\textbf {\bibinfo {volume} {2}},\ \bibinfo {pages}
  {79} (\bibinfo {year} {2018})}\BibitemShut {NoStop}%
\bibitem [{\citenamefont {Joseph}(2020)}]{Joseph2020}%
  \BibitemOpen
  \bibfield  {author} {\bibinfo {author} {\bibfnamefont {I.}~\bibnamefont
  {Joseph}},\ }\bibfield  {title} {\enquote {\bibinfo {title} {Koopman--von
  {N}eumann approach to quantum simulation of nonlinear classical dynamics},}\
  }\href {https://doi.org/10.1103/PhysRevResearch.2.043102} {\bibfield
  {journal} {\bibinfo  {journal} {Phys. Rev. Research}\ }\textbf {\bibinfo
  {volume} {2}},\ \bibinfo {pages} {043102} (\bibinfo {year}
  {2020})}\BibitemShut {NoStop}%
\bibitem [{\citenamefont {{Dodin}}\ and\ \citenamefont
  {{Startsev}}(2020)}]{Dodin2020}%
  \BibitemOpen
  \bibfield  {author} {\bibinfo {author} {\bibfnamefont {I.~Y.}\ \bibnamefont
  {{Dodin}}}\ and\ \bibinfo {author} {\bibfnamefont {E.~A.}\ \bibnamefont
  {{Startsev}}},\ }\bibfield  {title} {\enquote {\bibinfo {title} {{On
  applications of quantum computing to plasma simulations}},}\ }\href@noop {}
  {\bibfield  {journal} {\bibinfo  {journal} {arXiv e-prints}\ } (\bibinfo
  {year} {2020})},\ \Eprint {https://arxiv.org/abs/2005.14369}
  {arXiv:2005.14369} \BibitemShut {NoStop}%
\bibitem [{\citenamefont {Lubasch}\ \emph {et~al.}(2020)\citenamefont
  {Lubasch}, \citenamefont {Joo}, \citenamefont {Moinier}, \citenamefont
  {Kiffner},\ and\ \citenamefont {Jaksch}}]{Variational}%
  \BibitemOpen
  \bibfield  {author} {\bibinfo {author} {\bibfnamefont {M.}~\bibnamefont
  {Lubasch}}, \bibinfo {author} {\bibfnamefont {J.}~\bibnamefont {Joo}},
  \bibinfo {author} {\bibfnamefont {P.}~\bibnamefont {Moinier}}, \bibinfo
  {author} {\bibfnamefont {M.}~\bibnamefont {Kiffner}},\ and\ \bibinfo {author}
  {\bibfnamefont {D.}~\bibnamefont {Jaksch}},\ }\bibfield  {title} {\enquote
  {\bibinfo {title} {Variational quantum algorithms for nonlinear problems},}\
  }\href {https://doi.org/10.1103/PhysRevA.101.010301} {\bibfield  {journal}
  {\bibinfo  {journal} {Phys. Rev. A}\ }\textbf {\bibinfo {volume} {101}},\
  \bibinfo {pages} {010301} (\bibinfo {year} {2020})}\BibitemShut {NoStop}%
\bibitem [{\citenamefont {Steijl}(2020)}]{QuantumFluid}%
  \BibitemOpen
  \bibfield  {author} {\bibinfo {author} {\bibfnamefont {R.}~\bibnamefont
  {Steijl}},\ }\bibfield  {title} {\enquote {\bibinfo {title} {Quantum
  algorithms for nonlinear equations in fluid mechanics},}\ }\href
  {https://doi.org/10.5772/intechopen.95023} {\bibfield  {journal} {\bibinfo
  {journal} {IntechOpen}\ } (\bibinfo {year} {2020}),\
  10.5772/intechopen.95023}\BibitemShut {NoStop}%
\bibitem [{\citenamefont {{Liu}}\ \emph {et~al.}(2020)\citenamefont {{Liu}},
  \citenamefont {{{\O}ie Kolden}}, \citenamefont {{Krovi}}, \citenamefont
  {{Loureiro}}, \citenamefont {{Trivisa}},\ and\ \citenamefont
  {{Childs}}}]{Dissipative}%
  \BibitemOpen
  \bibfield  {author} {\bibinfo {author} {\bibfnamefont {J.-P.}\ \bibnamefont
  {{Liu}}}, \bibinfo {author} {\bibfnamefont {H.}~\bibnamefont {{{\O}ie
  Kolden}}}, \bibinfo {author} {\bibfnamefont {H.~K.}\ \bibnamefont {{Krovi}}},
  \bibinfo {author} {\bibfnamefont {N.~F.}\ \bibnamefont {{Loureiro}}},
  \bibinfo {author} {\bibfnamefont {K.}~\bibnamefont {{Trivisa}}},\ and\
  \bibinfo {author} {\bibfnamefont {A.~M.}\ \bibnamefont {{Childs}}},\
  }\bibfield  {title} {\enquote {\bibinfo {title} {Efficient quantum algorithm
  for dissipative nonlinear differential equations},}\ }\href@noop {}
  {\bibfield  {journal} {\bibinfo  {journal} {arXiv e-prints}\ } (\bibinfo
  {year} {2020})},\ \Eprint {https://arxiv.org/abs/2011.03185}
  {arXiv:2011.03185} \BibitemShut {NoStop}%
\bibitem [{\citenamefont {{Lloyd}}\ \emph {et~al.}(2020)\citenamefont
  {{Lloyd}}, \citenamefont {{De Palma}}, \citenamefont {{Gokler}},
  \citenamefont {{Kiani}}, \citenamefont {{Liu}}, \citenamefont {{Marvian}},
  \citenamefont {{Tennie}},\ and\ \citenamefont {{Palmer}}}]{Lloyd2020}%
  \BibitemOpen
  \bibfield  {author} {\bibinfo {author} {\bibfnamefont {S.}~\bibnamefont
  {{Lloyd}}}, \bibinfo {author} {\bibfnamefont {G.}~\bibnamefont {{De Palma}}},
  \bibinfo {author} {\bibfnamefont {C.}~\bibnamefont {{Gokler}}}, \bibinfo
  {author} {\bibfnamefont {B.}~\bibnamefont {{Kiani}}}, \bibinfo {author}
  {\bibfnamefont {Z.-W.}\ \bibnamefont {{Liu}}}, \bibinfo {author}
  {\bibfnamefont {M.}~\bibnamefont {{Marvian}}}, \bibinfo {author}
  {\bibfnamefont {F.}~\bibnamefont {{Tennie}}},\ and\ \bibinfo {author}
  {\bibfnamefont {T.}~\bibnamefont {{Palmer}}},\ }\bibfield  {title} {\enquote
  {\bibinfo {title} {Quantum algorithm for nonlinear differential equations},}\
  }\href@noop {} {\bibfield  {journal} {\bibinfo  {journal} {arXiv e-prints}\ }
  (\bibinfo {year} {2020})},\ \Eprint {https://arxiv.org/abs/2011.06571}
  {arXiv:2011.06571} \BibitemShut {NoStop}%
\bibitem [{\citenamefont {{Shi}}\ \emph {et~al.}(2020)\citenamefont {{Shi}},
  \citenamefont {{Castelli}}, \citenamefont {{Wu}}, \citenamefont {{Joseph}},
  \citenamefont {{Geyko}}, \citenamefont {{Graziani}}, \citenamefont {{Libby}},
  \citenamefont {{Parker}}, \citenamefont {{Rosen}}, \citenamefont
  {{Martinez}},\ and\ \citenamefont {{DuBois}}}]{Shi2020}%
  \BibitemOpen
  \bibfield  {author} {\bibinfo {author} {\bibfnamefont {Y.}~\bibnamefont
  {{Shi}}}, \bibinfo {author} {\bibfnamefont {A.~R.}\ \bibnamefont
  {{Castelli}}}, \bibinfo {author} {\bibfnamefont {X.}~\bibnamefont {{Wu}}},
  \bibinfo {author} {\bibfnamefont {I.}~\bibnamefont {{Joseph}}}, \bibinfo
  {author} {\bibfnamefont {V.}~\bibnamefont {{Geyko}}}, \bibinfo {author}
  {\bibfnamefont {F.~R.}\ \bibnamefont {{Graziani}}}, \bibinfo {author}
  {\bibfnamefont {S.~B.}\ \bibnamefont {{Libby}}}, \bibinfo {author}
  {\bibfnamefont {J.~B.}\ \bibnamefont {{Parker}}}, \bibinfo {author}
  {\bibfnamefont {Y.~J.}\ \bibnamefont {{Rosen}}}, \bibinfo {author}
  {\bibfnamefont {L.~A.}\ \bibnamefont {{Martinez}}},\ and\ \bibinfo {author}
  {\bibfnamefont {J.~L.}\ \bibnamefont {{DuBois}}},\ }\bibfield  {title}
  {\enquote {\bibinfo {title} {Simulating nonnative cubic interactions on noisy
  quantum machines},}\ }\href@noop {} {\bibfield  {journal} {\bibinfo
  {journal} {arXiv e-prints}\ } (\bibinfo {year} {2020})},\ \Eprint
  {https://arxiv.org/abs/2004.06885} {arXiv:2004.06885} \BibitemShut {NoStop}%
\bibitem [{\citenamefont {Feynman}(1982)}]{Feynman1982}%
  \BibitemOpen
  \bibfield  {author} {\bibinfo {author} {\bibfnamefont {R.~P.}\ \bibnamefont
  {Feynman}},\ }\bibfield  {title} {\enquote {\bibinfo {title} {Simulating
  physics with computers},}\ }\href {https://doi.org/10.1007/BF02650179}
  {\bibfield  {journal} {\bibinfo  {journal} {International Journal of
  Theoretical Physics}\ }\textbf {\bibinfo {volume} {21}},\ \bibinfo {pages}
  {467--488} (\bibinfo {year} {1982})}\BibitemShut {NoStop}%
\bibitem [{\citenamefont {Engel}, \citenamefont {Smith},\ and\ \citenamefont
  {Parker}(2019)}]{Engel2019}%
  \BibitemOpen
  \bibfield  {author} {\bibinfo {author} {\bibfnamefont {A.}~\bibnamefont
  {Engel}}, \bibinfo {author} {\bibfnamefont {G.}~\bibnamefont {Smith}},\ and\
  \bibinfo {author} {\bibfnamefont {S.~E.}\ \bibnamefont {Parker}},\ }\bibfield
   {title} {\enquote {\bibinfo {title} {Quantum algorithm for the {V}lasov
  equation},}\ }\href {https://doi.org/10.1103/PhysRevA.100.062315} {\bibfield
  {journal} {\bibinfo  {journal} {Phys. Rev. A}\ }\textbf {\bibinfo {volume}
  {100}},\ \bibinfo {pages} {062315} (\bibinfo {year} {2019})}\BibitemShut
  {NoStop}%
\bibitem [{\citenamefont {Parker}\ and\ \citenamefont
  {Joseph}(2020)}]{Parker2020}%
  \BibitemOpen
  \bibfield  {author} {\bibinfo {author} {\bibfnamefont {J.~B.}\ \bibnamefont
  {Parker}}\ and\ \bibinfo {author} {\bibfnamefont {I.}~\bibnamefont
  {Joseph}},\ }\bibfield  {title} {\enquote {\bibinfo {title} {Quantum phase
  estimation for a class of generalized eigenvalue problems},}\ }\href
  {https://doi.org/10.1103/PhysRevA.102.022422} {\bibfield  {journal} {\bibinfo
   {journal} {Phys. Rev. A}\ }\textbf {\bibinfo {volume} {102}},\ \bibinfo
  {pages} {022422} (\bibinfo {year} {2020})}\BibitemShut {NoStop}%
\bibitem [{\citenamefont {{Harrow}}, \citenamefont {{Hassidim}},\ and\
  \citenamefont {{Lloyd}}(2009)}]{Harrow2009}%
  \BibitemOpen
  \bibfield  {author} {\bibinfo {author} {\bibfnamefont {A.~W.}\ \bibnamefont
  {{Harrow}}}, \bibinfo {author} {\bibfnamefont {A.}~\bibnamefont
  {{Hassidim}}},\ and\ \bibinfo {author} {\bibfnamefont {S.}~\bibnamefont
  {{Lloyd}}},\ }\bibfield  {title} {\enquote {\bibinfo {title} {Quantum
  algorithm for linear systems of equations},}\ }\href
  {https://doi.org/10.1103/PhysRevLett.103.150502} {\bibfield  {journal}
  {\bibinfo  {journal} {Physical Review Letters}\ }\textbf {\bibinfo {volume}
  {103}},\ \bibinfo {eid} {150502} (\bibinfo {year} {2009})}\BibitemShut
  {NoStop}%
\bibitem [{\citenamefont {Childs}, \citenamefont {Kothari},\ and\ \citenamefont
  {Somma}(2017)}]{QLSA2017}%
  \BibitemOpen
  \bibfield  {author} {\bibinfo {author} {\bibfnamefont {A.}~\bibnamefont
  {Childs}}, \bibinfo {author} {\bibfnamefont {R.}~\bibnamefont {Kothari}},\
  and\ \bibinfo {author} {\bibfnamefont {R.}~\bibnamefont {Somma}},\ }\bibfield
   {title} {\enquote {\bibinfo {title} {Quantum algorithm for systems of linear
  equations with exponentially improved dependence on precision},}\ }\href
  {https://doi.org/10.1137/16M1087072} {\bibfield  {journal} {\bibinfo
  {journal} {SIAM Journal on Computing}\ }\textbf {\bibinfo {volume} {46}},\
  \bibinfo {pages} {1920--1950} (\bibinfo {year} {2017})}\BibitemShut {NoStop}%
\bibitem [{\citenamefont {Soklakov}\ and\ \citenamefont
  {Schack}(2006)}]{Soklakov2006}%
  \BibitemOpen
  \bibfield  {author} {\bibinfo {author} {\bibfnamefont {A.~N.}\ \bibnamefont
  {Soklakov}}\ and\ \bibinfo {author} {\bibfnamefont {R.}~\bibnamefont
  {Schack}},\ }\bibfield  {title} {\enquote {\bibinfo {title} {Efficient state
  preparation for a register of quantum bits},}\ }\href
  {https://doi.org/10.1103/PhysRevA.73.012307} {\bibfield  {journal} {\bibinfo
  {journal} {Phys. Rev. A}\ }\textbf {\bibinfo {volume} {73}},\ \bibinfo
  {pages} {012307} (\bibinfo {year} {2006})}\BibitemShut {NoStop}%
\bibitem [{\citenamefont {Berry}\ \emph {et~al.}(2017)\citenamefont {Berry},
  \citenamefont {Childs}, \citenamefont {Ostrander},\ and\ \citenamefont
  {Wang}}]{ODE2017}%
  \BibitemOpen
  \bibfield  {author} {\bibinfo {author} {\bibfnamefont {D.~W.}\ \bibnamefont
  {Berry}}, \bibinfo {author} {\bibfnamefont {A.~M.}\ \bibnamefont {Childs}},
  \bibinfo {author} {\bibfnamefont {A.}~\bibnamefont {Ostrander}},\ and\
  \bibinfo {author} {\bibfnamefont {G.}~\bibnamefont {Wang}},\ }\bibfield
  {title} {\enquote {\bibinfo {title} {Quantum algorithm for linear
  differential equations with exponentially improved dependence on
  precision},}\ }\href {https://doi.org/10.1007/s00220-017-3002-y} {\bibfield
  {journal} {\bibinfo  {journal} {Communications in Mathematical Physics}\
  }\textbf {\bibinfo {volume} {356}},\ \bibinfo {pages} {1057--1081} (\bibinfo
  {year} {2017})}\BibitemShut {NoStop}%
\bibitem [{\citenamefont {Kowalski}(1994{\natexlab{a}})}]{BlackBook1}%
  \BibitemOpen
  \bibfield  {author} {\bibinfo {author} {\bibfnamefont {K.}~\bibnamefont
  {Kowalski}},\ }\href {https://doi.org/10.1142/2345} {\emph {\bibinfo {title}
  {Methods of Hilbert Spaces in the Theory of Nonlinear Dynamical Systems}}}\
  (\bibinfo  {publisher} {World Scientific, Singapore},\ \bibinfo {year}
  {1994})\ p.~\bibinfo {pages} {1}\BibitemShut {NoStop}%
\bibitem [{\citenamefont {Koopman}(1931)}]{Koopman}%
  \BibitemOpen
  \bibfield  {author} {\bibinfo {author} {\bibfnamefont {B.~O.}\ \bibnamefont
  {Koopman}},\ }\bibfield  {title} {\enquote {\bibinfo {title} {Hamiltonian
  systems and transformation in {H}ilbert space},}\ }\href
  {https://doi.org/10.1073/pnas.17.5.315} {\bibfield  {journal} {\bibinfo
  {journal} {Proceedings of the National Academy of Sciences}\ }\textbf
  {\bibinfo {volume} {17}},\ \bibinfo {pages} {315--318} (\bibinfo {year}
  {1931})}\BibitemShut {NoStop}%
\bibitem [{\citenamefont {{von Neumann}}(1932{\natexlab{a}})}]{Neumann1}%
  \BibitemOpen
  \bibfield  {author} {\bibinfo {author} {\bibfnamefont {J.}~\bibnamefont {{von
  Neumann}}},\ }\bibfield  {title} {\enquote {\bibinfo {title} {Zur
  operatorenmethode in der klassischen mechanik},}\ }\href
  {http://www.jstor.org/stable/1968537} {\bibfield  {journal} {\bibinfo
  {journal} {Annals of Mathematics}\ }\textbf {\bibinfo {volume} {33}},\
  \bibinfo {pages} {587--642} (\bibinfo {year}
  {1932}{\natexlab{a}})}\BibitemShut {NoStop}%
\bibitem [{\citenamefont {{von Neumann}}(1932{\natexlab{b}})}]{Neumann2}%
  \BibitemOpen
  \bibfield  {author} {\bibinfo {author} {\bibfnamefont {J.}~\bibnamefont {{von
  Neumann}}},\ }\bibfield  {title} {\enquote {\bibinfo {title} {Zusatze zur
  arbeit ``zur operatorenmethode..."},}\ }\href
  {http://www.jstor.org/stable/1968225} {\bibfield  {journal} {\bibinfo
  {journal} {Annals of Mathematics}\ }\textbf {\bibinfo {volume} {33}},\
  \bibinfo {pages} {789--791} (\bibinfo {year}
  {1932}{\natexlab{b}})}\BibitemShut {NoStop}%
\bibitem [{\citenamefont {Carleman}(1932)}]{Carleman1932}%
  \BibitemOpen
  \bibfield  {author} {\bibinfo {author} {\bibfnamefont {T.}~\bibnamefont
  {Carleman}},\ }\bibfield  {title} {\enquote {\bibinfo {title} {Application de
  la théorie des équations intégrales linéaires aux systèmes d'équations
  différentielles non linéaires},}\ }\href
  {https://doi.org/10.1007/BF02546499} {\bibfield  {journal} {\bibinfo
  {journal} {Acta Math.}\ }\textbf {\bibinfo {volume} {59}},\ \bibinfo {pages}
  {63--87} (\bibinfo {year} {1932})}\BibitemShut {NoStop}%
\bibitem [{\citenamefont {Steeb}(1983)}]{Steeb1983}%
  \BibitemOpen
  \bibfield  {author} {\bibinfo {author} {\bibfnamefont {W.-H.}\ \bibnamefont
  {Steeb}},\ }\bibfield  {title} {\enquote {\bibinfo {title} {Embedding of
  nonlinear finite dimensional systems in linear infinite dimensional systems
  and {B}ose operators},}\ }\href@noop {} {\bibfield  {journal} {\bibinfo
  {journal} {Hadronic Journal}\ }\textbf {\bibinfo {volume} {6}},\ \bibinfo
  {pages} {68--76} (\bibinfo {year} {1983})}\BibitemShut {NoStop}%
\bibitem [{\citenamefont {Chirikov}, \citenamefont {Izrailev},\ and\
  \citenamefont {Shepelyansky}(1988)}]{Chirikov}%
  \BibitemOpen
  \bibfield  {author} {\bibinfo {author} {\bibfnamefont {B.}~\bibnamefont
  {Chirikov}}, \bibinfo {author} {\bibfnamefont {F.}~\bibnamefont {Izrailev}},\
  and\ \bibinfo {author} {\bibfnamefont {D.}~\bibnamefont {Shepelyansky}},\
  }\bibfield  {title} {\enquote {\bibinfo {title} {Quantum chaos: Localization
  vs. ergodicity},}\ }\href
  {https://doi.org/https://doi.org/10.1016/S0167-2789(98)90011-2} {\bibfield
  {journal} {\bibinfo  {journal} {Physica D: Nonlinear Phenomena}\ }\textbf
  {\bibinfo {volume} {33}},\ \bibinfo {pages} {77--88} (\bibinfo {year}
  {1988})}\BibitemShut {NoStop}%
\bibitem [{\citenamefont {Alanson}(1992)}]{Alanson1992}%
  \BibitemOpen
  \bibfield  {author} {\bibinfo {author} {\bibfnamefont {T.}~\bibnamefont
  {Alanson}},\ }\bibfield  {title} {\enquote {\bibinfo {title} {A “quantal”
  {H}ilbert space formulation for nonlinear dynamical systems in terms of
  probability amplitudes},}\ }\href
  {https://doi.org/https://doi.org/10.1016/0375-9601(92)90157-H} {\bibfield
  {journal} {\bibinfo  {journal} {Physics Letters A}\ }\textbf {\bibinfo
  {volume} {163}},\ \bibinfo {pages} {41--45} (\bibinfo {year}
  {1992})}\BibitemShut {NoStop}%
\bibitem [{\citenamefont {Kowalski}(1997)}]{Kow1997}%
  \BibitemOpen
  \bibfield  {author} {\bibinfo {author} {\bibfnamefont {K.}~\bibnamefont
  {Kowalski}},\ }\bibfield  {title} {\enquote {\bibinfo {title} {Nonlinear
  dynamical systems and classical orthogonal polynomials},}\ }\href
  {https://doi.org/10.1063/1.531990} {\bibfield  {journal} {\bibinfo  {journal}
  {Journal of Mathematical Physics}\ }\textbf {\bibinfo {volume} {38}},\
  \bibinfo {pages} {2483--2505} (\bibinfo {year} {1997})}\BibitemShut {NoStop}%
\bibitem [{\citenamefont {Kowalski}\ and\ \citenamefont
  {Steeb}(1991)}]{Kow1991}%
  \BibitemOpen
  \bibfield  {author} {\bibinfo {author} {\bibfnamefont {K.}~\bibnamefont
  {Kowalski}}\ and\ \bibinfo {author} {\bibfnamefont {W.-H.}\ \bibnamefont
  {Steeb}},\ }\href {https://doi.org/10.1142/1347} {\emph {\bibinfo {title}
  {Nonlinear Dynamical Systems and Carleman Linearization}}}\ (\bibinfo
  {publisher} {World Scientific, Singapore},\ \bibinfo {year}
  {1991})\BibitemShut {NoStop}%
\bibitem [{\citenamefont {Kowalski}(1994{\natexlab{b}})}]{Kow1994}%
  \BibitemOpen
  \bibfield  {author} {\bibinfo {author} {\bibfnamefont {K.}~\bibnamefont
  {Kowalski}},\ }\href {https://doi.org/10.1142/2345} {\emph {\bibinfo {title}
  {Methods of Hilbert Spaces in the Theory of Nonlinear Dynamical Systems}}}\
  (\bibinfo  {publisher} {World Scientific, Singapore},\ \bibinfo {year}
  {1994})\BibitemShut {NoStop}%
\bibitem [{\citenamefont {Varadarajan}(1970)}]{Varadarajan}%
  \BibitemOpen
  \bibfield  {author} {\bibinfo {author} {\bibfnamefont {V.~S.}\ \bibnamefont
  {Varadarajan}},\ }\href@noop {} {\emph {\bibinfo {title} {Geometry of Quantum
  Theory, Vol. II}}}\ (\bibinfo  {publisher} {Van Nostrand Reinhold, New
  York},\ \bibinfo {year} {1970})\BibitemShut {NoStop}%
\bibitem [{\citenamefont {Brassard}\ \emph {et~al.}(2002)\citenamefont
  {Brassard}, \citenamefont {H{\o}yer}, \citenamefont {Mosca},\ and\
  \citenamefont {Tapp}}]{Brassard2002}%
  \BibitemOpen
  \bibfield  {author} {\bibinfo {author} {\bibfnamefont {G.}~\bibnamefont
  {Brassard}}, \bibinfo {author} {\bibfnamefont {P.}~\bibnamefont {H{\o}yer}},
  \bibinfo {author} {\bibfnamefont {M.}~\bibnamefont {Mosca}},\ and\ \bibinfo
  {author} {\bibfnamefont {A.}~\bibnamefont {Tapp}},\ }\bibfield  {title}
  {\enquote {\bibinfo {title} {Quantum amplitude amplification and
  estimation},}\ }in\ \href {https://doi.org/10.1090/conm/305/05215} {\emph
  {\bibinfo {booktitle} {Quantum computation and information}}},\ \bibinfo
  {series} {Contemp. Math.}, Vol.\ \bibinfo {volume} {305}\ (\bibinfo
  {publisher} {Amer. Math. Soc., Providence, RI},\ \bibinfo {year} {2002})\
  pp.\ \bibinfo {pages} {53--74}\BibitemShut {NoStop}%
\bibitem [{\citenamefont {{Grover}}\ and\ \citenamefont
  {{Rudolph}}(2002)}]{Grover2002}%
  \BibitemOpen
  \bibfield  {author} {\bibinfo {author} {\bibfnamefont {L.}~\bibnamefont
  {{Grover}}}\ and\ \bibinfo {author} {\bibfnamefont {T.}~\bibnamefont
  {{Rudolph}}},\ }\bibfield  {title} {\enquote {\bibinfo {title} {Creating
  superpositions that correspond to efficiently integrable probability
  distributions},}\ }\href@noop {} {\bibfield  {journal} {\bibinfo  {journal}
  {arXiv e-prints}\ } (\bibinfo {year} {2002})},\ \Eprint
  {https://arxiv.org/abs/quant-ph/0208112} {arXiv:quant-ph/0208112}
  \BibitemShut {NoStop}%
\bibitem [{\citenamefont {Low}\ and\ \citenamefont {Chuang}(2017)}]{Low2017}%
  \BibitemOpen
  \bibfield  {author} {\bibinfo {author} {\bibfnamefont {G.~H.}\ \bibnamefont
  {Low}}\ and\ \bibinfo {author} {\bibfnamefont {I.~L.}\ \bibnamefont
  {Chuang}},\ }\bibfield  {title} {\enquote {\bibinfo {title} {Optimal
  {H}amiltonian simulation by quantum signal processing},}\ }\href
  {https://doi.org/10.1103/PhysRevLett.118.010501} {\bibfield  {journal}
  {\bibinfo  {journal} {Phys. Rev. Lett.}\ }\textbf {\bibinfo {volume} {118}},\
  \bibinfo {pages} {010501} (\bibinfo {year} {2017})}\BibitemShut {NoStop}%
\bibitem [{\citenamefont {Hoppensteadt}(2000)}]{Perturbations}%
  \BibitemOpen
  \bibfield  {author} {\bibinfo {author} {\bibfnamefont {F.~C.}\ \bibnamefont
  {Hoppensteadt}},\ }\href {https://doi.org/10.1007/b98824} {\emph {\bibinfo
  {title} {Analysis and Simulation of Chaotic Systems}}},\ \bibinfo {edition}
  {2nd}\ ed.\ (\bibinfo  {publisher} {Springer, New York},\ \bibinfo {year}
  {2000})\ pp.\ \bibinfo {pages} {152--153}\BibitemShut {NoStop}%
\bibitem [{\citenamefont {Low}\ and\ \citenamefont
  {Chuang}(2019)}]{Qubitization}%
  \BibitemOpen
  \bibfield  {author} {\bibinfo {author} {\bibfnamefont {G.~H.}\ \bibnamefont
  {Low}}\ and\ \bibinfo {author} {\bibfnamefont {I.~L.}\ \bibnamefont
  {Chuang}},\ }\bibfield  {title} {\enquote {\bibinfo {title} {Hamiltonian
  simulation by qubitization},}\ }\href
  {https://doi.org/10.22331/q-2019-07-12-163} {\bibfield  {journal} {\bibinfo
  {journal} {{Quantum}}\ }\textbf {\bibinfo {volume} {3}},\ \bibinfo {pages}
  {163} (\bibinfo {year} {2019})}\BibitemShut {NoStop}%
\bibitem [{\citenamefont {{Low}}\ and\ \citenamefont
  {{Chuang}}(2017)}]{LowSpectral}%
  \BibitemOpen
  \bibfield  {author} {\bibinfo {author} {\bibfnamefont {G.~H.}\ \bibnamefont
  {{Low}}}\ and\ \bibinfo {author} {\bibfnamefont {I.~L.}\ \bibnamefont
  {{Chuang}}},\ }\bibfield  {title} {\enquote {\bibinfo {title} {Hamiltonian
  simulation by uniform spectral amplification},}\ }\href@noop {} {\bibfield
  {journal} {\bibinfo  {journal} {arXiv e-prints}\ } (\bibinfo {year}
  {2017})},\ \Eprint {https://arxiv.org/abs/1707.05391} {arXiv:1707.05391}
  \BibitemShut {NoStop}%
\end{thebibliography}
\end{document}